\documentclass[reqno,12pt]{article}
\usepackage{amsfonts,amssymb,amsmath,amscd,bm,bbold,cite,delarray,subfigure}
\usepackage{xypic}
\usepackage[dvips]{color}
\usepackage[dvips]{graphicx}

\numberwithin{equation}{section}

\font\capital=rsfs12
\font\scriptcapital=rsfs10 at 7 truept
\font\scriptscriptcapital=rsfs10 at 5 truept
\def\scri{\fam=15}
\newcommand{\mathscri}[1]{{{\scri #1}}}

\font\sansserif=cmss12
\font\scriptsansserif=cmss12 at 7 truept
\font\scriptscriptsansserif=cmss10 at 5 truept
\font\euler=eusm10 at 12 truept
\font\scripteuler=eusm7
\font\scriptscripteuler=eusm5 
\newcommand{\im}{{\rm im}}

\begin{document}

\hrule\vskip.4cm
\hbox to 14.5 truecm{November 2006\hfil DFUB 06}
\hbox to 14.5 truecm{Version 1  \hfil hep-th/0611308}
\vskip.4cm\hrule
\vskip.7cm
\begin{large}
  \centerline{\textcolor{blue}{\bf BIHERMITIAN SUPERSYMMETRIC}}  
\centerline{\textcolor{blue}{\bf QUANTUM MECHANICS}}  
\end{large}
\vskip.2cm
\centerline{by}
\vskip.2cm
\centerline{\textcolor{blue}{\bf\bf Roberto Zucchini}}
\centerline{\it Dipartimento di Fisica, Universit\`a degli Studi di Bologna}
\centerline{\it V. Irnerio 46, I-40126 Bologna, Italy}
\centerline{\it I.N.F.N., sezione di Bologna, Italy}
\centerline{\it E--mail: zucchinir@bo.infn.it}
\vskip.7cm
\hrule
\vskip.7cm
\centerline{\textcolor{blue}{  \bf Abstract} }
\par\noindent
BiHermitian geometry, discovered long ago by Gates, Hull and Ro\v cek, is
the most general sigma model target space geometry allowing 
for $(2,2)$ world sheet supersymmetry. In this paper, we work out 
supersymmetric quantum  mechanics for a biHermitian target space. 
We display the full supersymmetry of the model and illustrate in detail 
its quantization procedure. Finally, we show that the quantized 
model reproduces the Hodge theory for compact twisted generalized Kaehler 
manifolds recently developed by Gualtieri in \cite{Gualtieri1}. This allows us to 
recover and put in a broader context the results on 
the biHermitian topological sigma models obtained by Kapustin and Li 
in \cite{Kapustin2}.

\vfill\eject

\begin{small}
\section{  \bf Introduction}
\label{sec:intro}
\end{small}

In a classic paper, Gates, Hull and Ro\v cek 
\cite{Gates} showed that, for a 2--dimensional sigma model, the most general
target space geometry allowing for $(2,2)$ supersymmetry 
was biHermitian or Kaehler with torsion geometry. This is characterized by 
a Riemannian metric $g_{ab}$, two generally non commuting complex structures 
$K_\pm{}^a{}_b$ and a closed $3$--form $H_{abc}$,
such that $g_{ab}$ is Hermitian with respect to both the $K_\pm{}^a{}_b$ and 
the $K_\pm{}^a{}_b$ are parallel with respect to two different metric
connections with torsion proportional to $\pm H_{abc}$ 
\cite{Rocek,Ivanov,Bogaerts,Lyakhovich}.

$(2,2)$ superconformal sigma models with Calabi--Yau target manifolds describe 
of compactifications of type II superstring and are therefore of considerable
physical interest. These are however nonlinear interacting field theories and,
so, are rather complicated and difficult to study. 
In 1988, Witten showed that a $(2,2)$ supersymmetric sigma model on a 
Calabi--Yau space could be twisted in two different ways,   
to yield the so called $A$ and $B$ topological sigma models \cite{Witten1,Witten2}.  
Unlike the original untwisted sigma model, the topological models are soluble
field theories: the calculation of observables can be carried out by standard 
methods of geometry and topology.
For the $A$ model, the ring of observables is found to be a deformation of
the complex de Rham cohomology $\bigoplus_p H^p(M,\mathbb{C})_{\rm qu}$,
going under the name of quantum cohomology, 
and all correlators can be shown to be symplectic invariants of $M$.
For the $B$ model, the ring of observables turns out to be isomorphic to 
$\bigoplus_{p,q} H^p(\wedge^qT^{1,0} M)$ and all
correlators are invariants of the complex structure on $M$.
Topological twisting of Calabi--Yau $(2,2)$ supersymmetric sigma models 
is therfore a very useful field theoretic procedure 
for the study of such field theories. 

Witten's analysis was restricted to the case where the sigma model target 
space geometry was Kaehler. This geometry is less general than that considered
by Gates, Hull and Ro\v cek, as it 
corresponds to the case where $K_+{}^a{}_b=\pm K_-{}^a{}_b$ and $H_{abc}=0$.
In the last few years, many attempts have been made to construct sigma models
with a biHermitian target manifolds, by invoking world sheet supersymmetry, 
employing the Batalin--Vilkovisky quantization algorithm, etc.
\cite{Kapustin1,Kapustin2,Kapustin3,Lindstrom2,Lindstrom3,
Lindstrom4,Lindstrom5,Chiantese,
Zabzine1,Zabzine2,Zucchini1,Zucchini2,Zucchini3,Zucchini4,Chuang, Li, Pestun1}.
A turning point in the quest towards accomplishing this goal was 
the realization that biHermitian geometry is naturally expressed in the language 
of generalized complex and Kaehler geometry
worked out by Hitchin and Gualtieri \cite{Hitchin1,Gualtieri,Zabzine3, Cavalcanti}. 

The subject of topological field theory itself can be traced back to Witten's
fundamental work on dynamical supersymmetry breaking \cite{Witten3,Witten4}.
Those findings led naturally to a reformulation of de Rham and Morse theory 
as supersymmetric quantum mechanics \cite{Witten5}. Since then, supersymmetric
quantum mechanics has been the object of intense study, for the rich relation
existing between the amount of 1--dimensional supersymmetry and 
the type of the differential geometric structure (Riemannian, Kaehler, 
hyperKaehler, etc.) present in target space (see e. g. \cite{Hull1} and
references therein). 

In this paper, we analyze supersymmetric quantum mechanics with a biHermitian 
target space. This model was first considered by Kapustin and Li in \cite{Kapustin2}
and was used to study the topological sector of $N=2$ sigma model with
$H$ flux. Here, we continue its study covering aspects not touched by  
Kapustin's and Li's analysis.  
We display the full supersymmetry of biHermitian supersymmetric 
quantum mechanics. Further, we illustrate in detail the quantization procedure
and show that demanding that the supersymmetry algebra is satisfied at the
quantum level solves all quantum ordering ambiguities.
Finally, we show that, upon quantization, the model reproduces the
Hodge theory for compact twisted generalized Kaehler manifolds recently developed
by Gualtieri \cite{Gualtieri1} (see also \cite{Cavalcanti1}), thereby
generalizing the well--known correspondence holding for ordinary Kaehler geometry.
We obtain in this way explicit local coordinate expressions of the 
relevant differential operators of Gualtieri's theory, which may be useful
in applications. We also explore the implications of our findings
for the geometrical interpretation of the 
biHermitian topological sigma models \cite{Zucchini4,Chuang}. In this way, we
recover and put in a broader context the results obtained by Kapustin and Li 
in \cite{Kapustin2}.

\vfill\eject

\vfill\eject

\begin{small}
\section{   \bf BiHermitian geometry}
\label{sec:biHermitian}
\end{small}

The target space geometry of the supersymmetric quantum mechanics studied in the
paper is biHermitian. Below, we review the basic facts of 
biHermitian geometry.

Let $M$ be a smooth manifold. A biHermitian structure $(g,H,K_\pm)$ on $M$ consists of
the following elements.
\par\noindent
$a$) A Riemannian metric $g_{ab}$ \footnote{\vphantom{$\Bigg]$} 
Here and below, indices are raised and lowered 
by using the metric $g_{ab}$.}.
\par\noindent
$b$) A closed 3--form $H_{abc}$  
\begin{equation}
\partial_{[a}H_{bcd]}=0.
\vphantom{\int}
\label{}
\end{equation}
\par\noindent
$c$) Two complex structures $K_\pm{}^a{}_b$,
\begin{align}
&K_\pm{}^a{}_cK_\pm{}^c{}_b=-\delta^a{}_b,
\vphantom{\int}
\label{}
\\
&K_\pm{}^d{}_a\partial_dK_\pm{}^c{}_b-K_\pm{}^d{}_b\partial_dK_\pm{}^c{}_a
-K_\pm{}^c{}_d\partial_aK_\pm{}^d{}_b+K_\pm{}^c{}_d\partial_bK_\pm{}^d{}_a=0.
\vphantom{\int}
\label{}
\end{align}
\par\noindent
They satisfy the following conditions. 
\par\noindent
$d$) $g_{ab}$ is Hermitian with respect to both $K_\pm{}^a{}_b$:
\begin{equation}
K_{\pm ab}+K_{\pm ba}=0.\vphantom{\int}
\label{}
\end{equation}
\par\noindent
$e$) 
The complex structures $K_\pm{}^a{}_b$ are parallel with respect to the
connections $\nabla_{\pm a}$ 
\begin{equation}
\nabla_{\pm a}K_\pm{}^b{}_c=0,
\label{}
\end{equation}
where the connection coefficients $\Gamma_\pm{}^a{}_{bc}$ are given by
\begin{equation}
\Gamma_\pm{}^a{}_{bc}=\Gamma^a{}_{bc}\pm\frac{1}{2}H^a{}_{bc}, 
\label{}
\end{equation}
$\Gamma^a{}_{bc}$ being the usual Levi--Civita connection coefficients.

The connections $\nabla_{\pm a}$ have a non vanishing torsion
$T_{\pm abc}$, which is totally antisymmetric and in fact equal to the 
3--form $H_{abc}$ up to sign,
\begin{equation}
T_{\pm abc}=\pm H_{abc}.
\label{}
\end{equation}
The Riemann tensors $R_{\pm abcd}$ of the $\nabla_{\pm a}$ satisfy a number of
relations, the most relevant of which are collected in appendix \ref{sec:appendixtens}. 

In biHermitian geometry, one is dealing with two generally non
commuting complex structures. As it turns out, it is not convenient 
to write the relevant tensor identities in the complex coordinates 
associated with either of them. General coordinates are 
definitely more natural and yield a more transparent formalism. 
Having this in mind, we define the complex tensors
\begin{equation}
\Lambda_{\pm}{}^a{}_b=\frac{1}{2}\big(\delta^a{}_b-iK_\pm{}^a{}_b\big).
\label{Pipm}
\end{equation}
The $\Lambda_{\pm}{}^a{}_b$ satisfy the relations 
\begin{subequations}\label{Pipm1}
\begin{align}
&\Lambda_{\pm}{}^a{}_c\Lambda_{\pm}{}^c{}_b=\Lambda_{\pm}{}^a{}_b,\vphantom{\int}
\label{Pipm1a}
\\
&\Lambda_{\pm}{}^a{}_b+\overline{\Lambda}{}_{\pm}{}^a{}_b=\delta^a{}_b,\vphantom{\int}
\label{Pipm1b}
\\
&\Lambda_{\pm}{}^a{}_b=\overline{\Lambda}{}_{\pm}{}_b{}^a.\vphantom{\int}
\label{Pipm1c}
\end{align}
\end{subequations}
Thus, the $\Lambda_{\pm}{}^a{}_b$ are projector valued endomorphisms of the
complexified tangent bundle $TM\otimes\mathbb{C}$.
The corresponding projection subbundles of $TM\otimes\mathbb{C}$ are the 
$K_\pm$--holomorphic
tangent bundles $T_\pm^{1,0}M$.

It turns out that the 3--form $H_{abc}$ is of type $(2,1)+(1,2)$ with respect
to both complex structures $K_\pm{}^a{}_b$,
\begin{equation}
H_{def}\Lambda_{\pm}{}^d{}_a\Lambda_{\pm}{}^e{}_b\Lambda_{\pm}{}^f{}_c=0
~~\text {and c.c.}
\label{}
\end{equation}
Other relations of the same type involving the Riemann tensors $R_{\pm abcd}$ are 
collected in appendix \ref{sec:appendixtens}. 

In \cite{Gualtieri}, Gualtieri has shown that biHermitian geometry is related 
to generalized Kaehler geometry. This, in turn, is part  
of generalized complex geometry. For a review of generalized complex and Kaehler 
geometry accessible to physicists, see \cite{Cavalcanti,Zabzine3}. 

\vfill\eject

\begin{small}
\section{  \bf The (2,2) supersymmetric sigma model}
\label{sec:(2,2)susysigma}
\end{small}

We shall review next the main properties of the biHermitian (2,2) supersymmetric 
sigma model, which are relevant in the following analysis. 

The base space of the model is a $1+1$ dimensional Minkoskian surface
$\Sigma$, usually taken to be a cylinder.
The target space of the model is a smooth manifold $M$ equipped with a
biHermtian structure $(g,H,K_\pm)$. 
The basic fields of the model are the embedding field $x^a$ of $\Sigma$ into $M$ 
and the spinor field $\psi_\pm{}^a$, which is valued in the vector bundle 
$x^*TM$ \footnote{\vphantom{$\bigg[$} Complying with an established use, 
here and in the following the indices $\pm$ are employed both to label the 
two complex structures $K_\pm$ of the relevant biHermitian structure and to denote
$2$--dimensional spinor indices. It should be clear from the context
what they stand for and no confusion should arise.}.

The action of biHermitian (2,2) supersymmetric 
sigma model is given by 
\begin{align}
S&=\int_\Sigma d^2\sigma\bigg[\frac{1}{2}(g_{ab}+b_{ab})(x)\partial_{++}x^a
\partial_{--}x^b
\label{22action}
\\
&\hskip1.9cm+\frac{i}{2}g_{ab}(x)(\psi_+{}^a\nabla_{+\,--}\psi_+{}^b
+\psi_-{}^a\nabla_{-\,++}\psi_-{}^b)
\nonumber\\
&\hskip1.9cm
+\frac{1}{4}R_{+abcd}(x)\psi_+{}^a\psi_+{}^b\psi_-{}^c\psi_-{}^d
\bigg],
\nonumber
\end{align}
where $\partial_{\pm\pm}=\partial_0\pm\partial_1$, \hphantom{xxxxxxxxxxx} 
\begin{equation}
\nabla_{\pm\,\mp\mp}=\partial_{\mp\mp}+\Gamma_\pm{}^\cdot{}_{c\,\cdot}(x)\partial_{\mp\mp}x^c
\end{equation}
and the field $b_{ab}$ is related to $H_{abc}$ as  
\begin{equation}\label{H=db}
H_{abc}=\partial_ab_{bc}+\partial_bb_{ca}+\partial_cb_{ab}.
\end{equation}

The $(2,2)$ supersymmetry variations of the basic fields can be written in several ways. 
We shall write  them in the following convenient form
\begin{subequations}\label{22vars}
\begin{align}
\delta_S x^a&=
i\big[\alpha^+\Lambda_+{}^a{}_b(x)\psi_+{}^b
+\tilde\alpha^+\overline{\Lambda}_+{}^a{}_b(x)\psi_+{}^b
\label{22varsa}\\
&\hphantom{=}~~~~+\alpha^-\Lambda_-{}^a{}_b(x)\psi_-{}^b
+\tilde\alpha^-\overline{\Lambda}_-{}^a{}_b(x)\psi_-{}^b\big],~~~~~~~~~~~~~~~~~
\nonumber
\end{align}
\pagebreak[2]
\begin{align}
\delta_S\psi_\pm{}^a&=-\alpha^\pm\overline{\Lambda}_\pm{}^a{}_b(x)\partial_{\pm\pm}x^b
-\tilde\alpha^\pm\Lambda_\pm{}^a{}_b(x)\partial_{\pm\pm}x^b
\label{22varsb}\\
&\hphantom{=}\,-i\Gamma_\pm{}^a{}_{bc}(x)\big[\alpha^+\Lambda_+{}^b{}_d(x)\psi_+{}^d
+\tilde\alpha^+\overline{\Lambda}_+{}^b{}_d(x)\psi_+{}^d
\nonumber\\
&\hphantom{\,=-i\Gamma_\pm{}^a{}_{bc}(x)\Big[}+\alpha^-\Lambda_-{}^b{}_d(x)\psi_-{}^d
+\tilde\alpha^-\overline{\Lambda}_-{}^b{}_d(x)\psi_-{}^d\big]\psi_\pm{}^c
\nonumber\\
&\hphantom{=}\,\pm i H^a{}_{bc}(x)\big[
\alpha^\pm\Lambda_\pm{}^b{}_d(x)\psi_\pm{}^d
+\tilde\alpha^\pm\overline{\Lambda}_\pm{}^b{}_d(x)\psi_\pm{}^d\big]\psi_\pm{}^c
\nonumber\\
&\hphantom{=}\,\mp\frac{i}{2}
\big(
\alpha^\pm\Lambda_\pm{}^a{}_d
+\tilde\alpha^\pm\overline{\Lambda}_\pm{}^a{}_d\big)H^d{}_{bc}(x)
\psi_\pm{}^b\psi_\pm{}^c,
\nonumber
\end{align}
\end{subequations}
where $\alpha^\pm$, $\tilde\alpha^\pm$ are constant Grassmann parameters. 
$\delta_S$ generates a $(2,2)$ supersymmetry algebra on shell.
The action $S$ enjoys $(2,2)$ supersymmetry, so that
\begin{equation}
\delta_S S=0.
\end{equation}

The biHermitian $(2,2)$ supersymmetric sigma model is characterized also by 
two types of $R$ symmetry: the $U(1)_V$ vector and the $U(1)_A$ axial $R$ symmetries
\begin{subequations}\label{Rvars}
\begin{align}
\delta_Rx^a&=0,
\label{Rvarsa}\\
\delta_R\psi_\pm{}^a&=i(\epsilon_V\pm \epsilon_A)\Lambda_\pm{}^a{}_b(x)\psi_\pm{}^b
-i(\epsilon_V\pm \epsilon_A)\overline{\Lambda}_\pm{}^a{}_b(x)\psi_\pm{}^b,\label{Rvarsb}
\end{align}
\end{subequations}
where $\epsilon_V$, $\epsilon_A$ are infinitesimal real parameters. 
Classically, the action $S$ enjoys both types of $R$ symmetry, so that 
\begin{equation}
\delta_RS=0.
\end{equation}
As is well known, at the quantum level, the $R$ symmetries are spoiled by
anomalies unless certain topological conditions on the
target manifold $M$ are satisfied \cite{Kapustin2}.

It is convenient to introduce the projected spinor fields
\begin{subequations}\label{Lambdapm}
\begin{align}
\chi_\pm{}^a&=\Lambda_\pm{}^a{}_b(x)\psi_\pm{}^b,\\
\overline{\chi}{}_\pm{}^a&=\overline{\Lambda}{}_\pm{}^a{}_b(x)\psi_\pm{}^b.
\end{align}
\end{subequations} 
In terms of these, the action $S$ reads 
\begin{align}
S&=\int_\Sigma d^2\sigma\bigg[\frac{1}{2}(g_{ab}+b_{ab})(x)\partial_{++}x^a
\partial_{--}x^b
\label{22actionpr}\\
&\hskip1.9cm+ig_{ab}(x)(\overline{\chi}{}_+{}^a\nabla_{+\,--}\chi_+{}^b
+\overline{\chi}{}_-{}^a\nabla_{-\,++}\chi_-{}^b)
\nonumber\\
&\hskip1.9cm
+R_{+abcd}(x)\overline{\chi}_+{}^a\chi_+{}^b\overline{\chi}{}_-{}^c\chi_-{}^d
\bigg].
\nonumber
\end{align}
The $(2,2)$ supersymmetry variations \eqref{22vars} take the simpler form
\begin{subequations}\label{22varspr}
\begin{align}
\delta_S x^a&=
i\big[\alpha^+\chi_+{}^a+\tilde\alpha^+\overline{\chi}{}_+{}^a
+\alpha^-\chi_-{}^a+\tilde\alpha^-\overline{\chi}{}_-{}^a\big],
\label{22varspra}\\
\delta_S\chi_\pm{}^a&=-i\Gamma_\pm{}^a{}_{bc}(x)
\big[\alpha^+\chi_+{}^b+\tilde\alpha^+\overline{\chi}{}_+{}^b
+\alpha^-\chi_-{}^b+\tilde\alpha^-\overline{\chi}{}_-{}^b\big]\chi_\pm{}^c
\label{22varsprb}\\
&\hphantom{~~~}
\pm\frac{i}{2}\alpha^\pm H^a{}_{bc}(x)\chi_\pm{}^b\chi_\pm{}^c
-\tilde\alpha^\pm\Lambda_\pm{}^a{}_b(x)\big[\partial_{\pm\pm}x^b
\mp iH^b{}_{cd}(x)\overline{\chi}{}_\pm{}^c\chi_\pm{}^d\big],
\nonumber\\
\delta_S\overline{\chi}{}_\pm{}^a&=-i\Gamma_\pm{}^a{}_{bc}(x)
\big[\alpha^+\chi_+{}^b+\tilde\alpha^+\overline{\chi}{}_+{}^b
+\alpha^-\chi_-{}^b+\tilde\alpha^-\overline{\chi}{}_-{}^b\big]\overline{\chi}{}_\pm{}^c
\label{22varsprc}\\
&\hphantom{~~~}
\pm\frac{i}{2}\tilde\alpha^\pm H^a{}_{bc}(x)\overline{\chi}{}_\pm{}^b\overline{\chi}{}_\pm{}^c
-\alpha^\pm\overline{\Lambda}{}_\pm{}^a{}_b(x)\big[\partial_{\pm\pm}x^b
\mp iH^b{}_{cd}(x)\chi_\pm{}^c\overline{\chi}{}_\pm{}^d\big].\nonumber
\end{align}
\end{subequations}
Similarly, the $R$ symmetry \eqref{Rvars} can be cast in simple form as
\begin{subequations}\label{Rvarspr}
\begin{align}
\delta_Rx^a&=0,
\label{Rvarspra}\\
\delta_R\chi_\pm{}^a&=+i(\epsilon_V\pm \epsilon_A)\chi_\pm{}^a,
\label{Rvarsprb}\\
\delta_R\overline{\chi}{}_\pm{}^a&=-i(\epsilon_V\pm \epsilon_A)
\overline{\chi}{}_\pm{}^a.
\label{Rvarsprc}
\end{align}
\end{subequations}
This projected spinor formulation of the $(2,2)$ supersymmetric sigma model turns out to be 
far more convenient in the following analysis than the more conventional one
reviewed in the first half of this section.

\vfill\eject

\begin{small}
\section{   \bf The biHermitian susy quantum mechanics}
\label{sec:susyqm}
\end{small}

We can obtain the biHermitian supersymmetric quantum mechanics
from the biHermitian $(2,2)$ supersymmetric sigma model by taking the world sheet
$\Sigma$ to be of the form $\Sigma=T\times S^1$ with $T=\mathbb{R}$ and dimensionally 
reduce form $1+1$ to $1+0$ by shrinking the $S^1$ factor to a point.

We use the projected spinor formalism illustrated in 
section \ref{sec:(2,2)susysigma}.
Then, the  biHermitian supersymmetric quantum mechanics action $S_{QM}$ reads 
\begin{align}
S_{QM}&=\int_T dt\bigg[\frac{1}{2}g_{ab}(x)\partial_tx^a
\partial_tx^b+ig_{ab}(x)(\overline{\chi}{}_+{}^a\nabla_{+t}\chi_+{}^b
+\overline{\chi}{}_-{}^a\nabla_{-t}\chi_-{}^b)
\label{qmactionpr}\\
&\hskip7.4cm
+R_{+abcd}(x)\overline{\chi}_+{}^a\chi_+{}^b\overline{\chi}{}_-{}^c\chi_-{}^d
\bigg],
\nonumber
\end{align}
where the nabla operator $\nabla_{\pm t}$ is given by 
\begin{equation}
\nabla_{\pm t}=\partial_t+\Gamma_\pm{}^\cdot{}_{c\,\cdot}(x)\partial_tx^c.
\end{equation}
The $b$ field no longer appears in the action, as is obvious from dimensional
considerations. Note that, by \eqref{Lambdapm}, the fermionic variables 
$\chi_\pm{}^a$, $\overline{\chi}{}_\pm{}^a$ are constrained: 
$\overline{\Lambda}_\pm{}^a{}_b(x)\chi_\pm{}^b=
\Lambda_\pm{}^a{}_b(x)\overline{\chi}{}_\pm{}^b=0$.

The supersymmetry variations are easily read off from \eqref{22varspr}.
It is convenient to decompose supersymmetry variation operator $\delta_S$ as  
\begin{equation}\label{qpm}
\delta_S=\alpha^+q_++\tilde\alpha^+\overline{q}{}_+
+\alpha^-q_-+\tilde\alpha^-\overline{q}{}_-,
\end{equation}
where the fermionic variation operators $q_\pm $, $\overline{q}{}_\pm$ are given by
\begin{subequations}\label{qmvarspr}
\begin{align}
q_\pm x^a&=i\chi_\pm{}^a,
\\
\overline{q}{}_\pm x^a&=i\overline{\chi}{}_\pm{}^a,
\\
q_\pm\chi_\pm{}^a&=0,
\\
q_\mp\chi_\pm{}^a&=-i\Gamma_\pm{}^a{}_{bc}(x)\chi_\mp{}^b\chi_\pm{}^c,
\\
\overline{q}{}_\pm\chi_\pm{}^a&=-i\Gamma_\pm{}^a{}_{bc}(x)\overline{\chi}{}_\pm{}^b\chi_\pm{}^c
-\Lambda_\pm{}^a{}_b(x)\big[\partial_tx^b
\mp i H^b{}_{cd}(x)\overline{\chi}{}_\pm{}^c\chi_\pm{}^d\big],\\
\overline{q}{}_\mp\chi_\pm{}^a&=-i\Gamma_\pm{}^a{}_{bc}(x)\overline{\chi}{}_\mp{}^b\chi_\pm{}^c,
\end{align}
\pagebreak[2]
\begin{align}
q_\pm\overline{\chi}{}_\pm{}^a&=
-i\Gamma_\pm{}^a{}_{bc}(x)\chi_\pm{}^b\overline{\chi}{}_\pm{}^c
-\overline{\Lambda}_\pm{}^a{}_b(x)\big[\partial_tx^b
\mp i H^b{}_{cd}(x)\chi_\pm{}^c\overline{\chi}{}_\pm{}^d\big],
\\
q_\mp\overline{\chi}{}_\pm{}^a&=-i\Gamma_\pm{}^a{}_{bc}(x)\chi_\mp{}^b\overline{\chi}{}_\pm{}^c,
\\
\overline{q}{}_\pm\overline{\chi}{}_\pm{}^a&=0,
\\
\overline{q}{}_\mp\overline{\chi}{}_\pm{}^a&=
-i\Gamma_\pm{}^a{}_{bc}(x)\overline{\chi}{}_\mp{}^b\overline{\chi}{}_\pm{}^c.
\end{align}
\end{subequations}

The $(2,2)$ supersymmetry of the sigma model action $S$ is inherited by the quantum
mechanics action $S_{QM}$, so that
\begin{equation}
q_\pm S_{QM}=\overline{q}{}_\pm S_{QM}=0.
\end{equation}
The associated four conserved supercharges $Q_{\pm}$, $\overline{Q}{}_\pm$ can be
computed by the Noether procedure by letting the supersymmetry parameters
$\alpha^\pm$, $\tilde\alpha^\pm$ to be time dependent:
\begin{equation}
\delta_S S_{QM}=\int_T dt \,i\big[\partial_t\alpha^+Q_+
+\partial_t\tilde\alpha^+\overline{Q}{}_+
+\partial_t\alpha^-Q_-
+\partial_t\tilde\alpha^-\overline{Q}{}_-\big].
\end{equation}
In this way, one finds that
\begin{subequations}\label{clsusych}
\begin{align}
Q_\pm&=g_{ab}(x)\chi_\pm{}^a\partial_tx^b
\mp\frac{i}{2}H_{abc}(x)\chi_\pm{}^a\chi_\pm{}^b\overline{\chi}{}_\pm{}^c,
\label{clsusycha}\\
\overline{Q}{}_\pm&=g_{ab}(x)\overline{\chi}{}_\pm{}^a\partial_tx^b
\mp\frac{i}{2}H_{abc}(x)\overline{\chi}{}_\pm{}^a\overline{\chi}{}_\pm{}^b\chi_\pm{}^c.
\label{clsusychb}
\end{align}
\end{subequations}

Similarly, the $R$ symmetry variations can be read off from \eqref{Rvarspr}.
It is convenient to decompose $R$ variation operator $\delta_R$ as 
\begin{equation}\label{fvfa}
\delta_R=i\big(\epsilon_Vf_V+\epsilon_Af_A\big),
\end{equation}
where the bosonic variation operators $f_V$, $f_a$ are given by
\begin{subequations}\label{Rqmvarspr}
\begin{align}
f_Vx^a&=0,
\\
f_Ax^a&=0,
\\
f_V\chi_\pm{}^a&=+\chi_\pm{}^a,
\\
f_A\chi_\pm{}^a&=\pm\chi_\pm{}^a,
\\
f_V\overline{\chi}{}_\pm{}^a&=-\overline{\chi}{}_\pm{}^a,
\\
f_A\overline{\chi}{}_\pm{}^a&=\mp\overline{\chi}{}_\pm{}^a.
\end{align}
\end{subequations}
The $R$ symmetry of the sigma model action $S$ is inherited by the quantum
mechanics action $S_{QM}$, so that one has
\begin{equation}
f_VS_{QM}=f_AS_{QM}=0.
\end{equation}
The associated two conserved $R$ charges $F_V$, $F_A$ can be
computed easily again by the Noether procedure by letting the $R$ Symmetry parameters
$\epsilon_V$, $\epsilon_A$ to be time dependent: \hphantom{xxxxxxxxxxxxxxxxxxxxxx}
\begin{equation}
\delta_R S_{QM}=-\int_T dt \big[\partial_t\epsilon_VF_V
+\partial_t\epsilon_AF_A\big]. 
\end{equation}
In this way, one finds that 
\begin{subequations}\label{clRch}
\begin{align}
F_V=g_{ab}(x)\big(\overline{\chi}{}_+{}^a\chi_+{}^b+\overline{\chi}{}_-{}^a\chi_-{}^b\big),
\label{clRcha}\\
F_A=g_{ab}(x)\big(\overline{\chi}{}_+{}^a\chi_+{}^b-\overline{\chi}{}_-{}^a\chi_-{}^b\big).
\label{clRchb}
\end{align}
\end{subequations}

It is straightforward to see that 
\begin{subequations}\label{clsymalg}
\begin{align}
&q_\pm{}^2\approx 0,\\
&q_\pm q_\mp+q_\mp q_\pm\approx 0,\\
&\overline{q}{}_\pm{}^2\approx 0,\\
&\overline{q}{}_\pm\overline{q}{}_\mp+\overline{q}{}_\mp\overline{q}{}_\pm\approx 0,\\
&q_\pm\overline{q}{}_\pm+\overline{q}{}_\pm q_\pm\approx -i\partial_t,\\
&q_\pm\overline{q}{}_\mp+\overline{q}{}_\mp q_\pm\approx 0,\\
&f_Vq_\pm-q_\pm f_V \approx +q_\pm,\\
&f_Aq_\pm-q_\pm f_A \approx \pm q_\pm,\\
&f_V\overline{q}{}_\pm-\overline{q}{}_\pm f_V\approx -\overline{q}{}_\pm,\\
&f_A\overline{q}{}_\pm-\overline{q}{}_\pm f_A\approx \mp \overline{q}{}_\pm, \\
&f_Vf_A-f_Af_V\approx 0.
\end{align}
\end{subequations}
where $\approx$ denotes equality on shell. In more precise terms, this means
the following. Let $\mathscri{F}$ denote the algebra of local composite 
fields and let $\mathscri{E}$ be the ideal of $\mathscri{F}$ generated 
by the field equations. Then, it can be shown that $q_{\pm}$,
$\overline{q}{}_\pm$, $f_V$, $f_A$ leave $\mathscri{E}$ invariant.  
In this way, $q_{\pm}$, $\overline{q}{}_\pm$, $f_V$, $f_A$ define variations
operators on the quotient algebra $\mathscri{F}_\mathscri{E}=\mathscri{F}/\mathscri{E}$. 
As such, they satisfy the algebra \eqref{clsymalg} with the on shell equality
sign $\approx$ replaced by the usual equality sign $=$. 

The above analysis shows that biHermitian supersymmetric quantum mechanics
enjoys a $N=4$ supersymmetry. If the target space is endowed with a structure
containing the given biHermitian structure as a substructure, e. g. a hyperKaehler 
structure, the amount of supersymmetry may be enhanced \cite{Hull1}. 
We shall not explore this possibility in this paper.

\vfill\eject

\begin{small}
\section{   \bf Quantization}
\label{sec:canquant}
\end{small}

From the expression of the action $S_{QM}$ of biHermitian supersymmetric 
quantum mechanics, eq. \eqref{qmactionpr}, one can easily read off the
classical Lagrangian 
\begin{align}
L_{QM}&=\frac{1}{2}g_{ab}(x)\partial_tx^a\partial_tx^b
+\frac{i}{2}g_{ab}(x)\big(\overline{\chi}{}_+{}^a\nabla_{+t}\chi_+{}^b
-\nabla_{+t}\overline{\chi}{}_+{}^a\chi_+{}^b
\label{qmlagrpr}\\
&\hskip2.3cm+\overline{\chi}{}_-{}^a\nabla_{-t}\chi_-{}^b
-\nabla_{-t}\overline{\chi}{}_-{}^a\chi_-{}^b\big)
+R_{+abcd}(x)\overline{\chi}_+{}^a\chi_+{}^b\overline{\chi}{}_-{}^c\chi_-{}^d.
\nonumber
\end{align}
In order $L_{QM}$ to be real, the kinetic term of the fermion coordinates 
$\chi_\pm{}^a$, $\overline{\chi}{}_\pm{}^a$ in \eqref{qmactionpr} has been cast 
in symmetric form by adding a 
total time derivative term. 

The conjugate momenta of the coordinates $x^a$, $\chi_\pm{}^a$,
$\overline{\chi}{}_\pm{}^a$ are defined 
\begin{subequations}
\begin{align}
\pi_a&=\frac{\partial L_{QM}}{\partial\partial_t x^a},\\
\lambda_{\pm a}
&=-\frac{\partial L_{QM}}{\partial\nabla_{\pm t}\chi_\pm{}^a},\\
\overline{\lambda}{}_{\pm a}
&=+\frac{\partial L_{QM}}{\partial\nabla_{\pm t}\overline{\chi}{}_\pm{}^a}.
\end{align}
\end{subequations}
To have manifest covariance, we define the fermionic momenta
by differentiating with respect to $\nabla_{\pm t}\chi_\pm{}^a$,
$\nabla_{\pm t}\overline{\chi}{}_\pm{}^a$ rather than 
$\partial_t \chi_\pm{}^a$, $\partial_t \overline{\chi}{}_\pm{}^a$. 
In this way, the implicit dependence of $\nabla_{\pm t}\chi_\pm{}^a$,
$\nabla_{\pm t}\overline{\chi}{}_\pm{}^a$ on $\partial_t x^a$ is disregarded
in the computation of $\pi_a$. Therefore, the momenta $\pi_a$ are not canonical.
The advantages of this way of proceeding will become clear in due course. 
Explicitly, one has 
\begin{subequations}\label{momexplexpr}
\begin{align}
\pi_a&=g_{ab}(x)\partial_tx^b,
\label{momexplexpra}\\
\lambda_{\pm a}&=\frac{i}{2}g_{ab}(x)\overline{\chi}{}_\pm{}^b,
\label{momexplexprb}\\
\overline{\lambda}{}_{\pm a}&=-\frac{i}{2}g_{ab}(x)\chi_\pm{}^b.
\label{momexplexprc}
\end{align}
\end{subequations}
Note that the constraints $\overline{\Lambda}_\pm{}^a{}_b(x)\chi_\pm{}^b=
\Lambda_\pm{}^a{}_b(x)\overline{\chi}{}_\pm{}^b=0$ imply correspondingly the
constraints $\overline{\Lambda}_\pm{}^b{}_a(x)\lambda_{\pm b}=
\Lambda_\pm{}^b{}_a(x)\overline{\lambda}{}_{\pm b}=0$.

The classical Hamiltonian is computed as usual
\begin{align}
H_{QM}=\pi_a\partial_tx^a+\lambda_{+a}\nabla_{+t}\chi_+{}^a&+\lambda_{-a}\nabla_{-t}\chi_-{}^a
\label{qmhamiltpr}\\
&-\overline{\lambda}{}_{+a}\nabla_{+t}\overline{\chi}{}_+{}^a
-\overline{\lambda}{}_{-a}\nabla_{-t}\overline{\chi}{}_-{}^a-L_{QM}.
\nonumber
\end{align}
The resulting expression of $H_{QM}$ is 
\begin{equation}
H_{QM}=\frac{1}{2}g^{ab}(x)\pi_a\pi_b
-R_{+abcd}(x)\overline{\chi}_+{}^a\chi_+{}^b\overline{\chi}{}_-{}^c\chi_-{}^d.
\end{equation}

The graded Poisson brackets of the coordinates $x^a$, $\chi_\pm{}^a$,
$\overline{\chi}{}_\pm{}^a$ and momenta $\pi_a$, $\lambda_{\pm a}$,
$\overline{\lambda}{}_{\pm a}$ are given by
\begin{subequations}\label{Poisbrkts}
\begin{align}
\{x^a,\pi_b\}_P&=\delta^a{}_b, 
\label{Poisbrktsa}\\
\{\pi_a,\pi_b\}_P&=
R_+{}^c{}_{dab}(x)(\lambda_{+c}\chi_+{}^d-\overline{\lambda}{}_{+c}\overline{\chi}{}_+{}^d)
\label{Poisbrktsb}\\
&\hskip4cm
+R_-{}^c{}_{dab}(x)(\lambda_{-c}\chi_-{}^d-\overline{\lambda}{}_{-c}\overline{\chi}{}_-{}^d),
\nonumber\\
\{\pi_a,\chi_\pm{}^b\}_P&=\Gamma_{\pm}{}^b{}_{ac}(x)\chi_\pm{}^c,
\label{Poisbrktsc}\\
\{\pi_a,\overline{\chi}{}_\pm{}^b\}_P&=\Gamma_{\pm}{}^b{}_{ac}(x)\overline{\chi}{}_\pm{}^c,
\label{Poisbrktsd}\\
\{\pi_a,\lambda_{\pm b}\}_P&=-\Gamma_{\pm}{}^c{}_{ab}(x)\lambda_{\pm c},
\label{Poisbrktse}\\
\{\pi_a,\overline{\lambda}{}_{\pm b}\}_P&=-\Gamma_{\pm}{}^c{}_{ab}(x)\overline{\lambda}{}_{\pm c},
\label{Poisbrktsf}\\
\{\chi_\pm{}^a,\lambda_{\pm b}\}_P&=\Lambda_\pm{}^a{}_b(x),
\label{Poisbrktsg}\\
\{\overline{\chi}{}_\pm{}^a,\overline{\lambda}{}_{\pm b}\}_P&=
-\overline{\Lambda}_\pm{}^a{}_b(x),
\label{Poisbrktsh}
\end{align}
\end{subequations}
all remaining Poisson brackets vanishing identically. 
The form of the brackets \eqref{Poisbrktsc}--\eqref{Poisbrktsf} is dictated by
covariance and the constraints $\overline{\Lambda}_\pm{}^a{}_b(x)\chi_\pm{}^b=
\Lambda_\pm{}^a{}_b(x)\overline{\chi}{}_\pm{}^b=0$, 
$\overline{\Lambda}_\pm{}^b{}_a(x)\lambda_{\pm b}=
\Lambda_\pm{}^b{}_a(x)\overline{\lambda}{}_{\pm b}=0$.
The form of the Poisson bracket \eqref{Poisbrktsb} is then essentially
determined by the fulfillment of the Jacobi identity.

From \eqref{momexplexprb}, \eqref{momexplexprc}, it follows that the
constraints 
\begin{subequations}\label{constraints}
\begin{align}
C_{\pm a}&:=\lambda_{\pm
  a}-\frac{i}{2}g_{ab}(x)\overline{\chi}{}_\pm{}^b\simeq 0,
\label{constraintsa}\\
\overline{C}{}_{\pm a}&:=\overline{\lambda}{}_{\pm a}+\frac{i}{2}g_{ab}(x)\chi_\pm{}^b\simeq 0
\label{constraintsb}
\end{align}
\end{subequations}
hold, where $\simeq$ denotes weak equality in Dirac's sense. These constraints are second class,
as follows from the Poisson brackets
\begin{equation}\label{IIclass}
\{C_{\pm a},\overline{C}{}_{\pm b}\}_P=i\Lambda_{\pm ab}(x),
\end{equation}
all remaining Poisson brackets of the constraints vanishing.
The resulting graded Dirac brackets of the independent variables $x^a$, $\pi_a$, 
$\chi_\pm{}^a$, $\overline{\chi}{}_\pm{}^a$ are easily computed:
\pagebreak[2]
\begin{subequations}\label{Diracbrkts}
\begin{align}
\{x^a,\pi_b\}_D&=\delta^a{}_b, 
\label{Diracbrktsa}\\
\{\pi_a,\pi_b\}_D&=
iR_{+cdab}(x)\overline{\chi}{}_+{}^c\chi_+{}^d
+iR_{-cdab}(x)\overline{\chi}{}_-{}^c\chi_-{}^d,
\label{Diracbrktsb}\\
\{\pi_a,\chi_\pm{}^b\}_D&=\Gamma_{\pm}{}^b{}_{ac}(x)\chi_\pm{}^c,
\label{Diracbrktsc}\\
\{\pi_a,\overline{\chi}{}_\pm{}^b\}_D&=\Gamma_{\pm}{}^b{}_{ac}(x)\overline{\chi}{}_\pm{}^c,
\label{Diracbrktsd}\\
\{\chi_\pm{}^a,\overline{\chi}{}_\pm{}^b\}_D&=-i\Lambda_\pm{}^{ab}(x),
\label{Diracbrktse}
\end{align}
\end{subequations}
all remaining Dirac brackets vanishing identically. 

To quantize the theory, we promote the variables $x^a$, $\pi_a$, 
$\chi_\pm{}^a$, $\overline{\chi}{}_\pm{}^a$ to operators
and stipulate that their graded commutators are  
given by the formal substitution $\{\,,\,\}_D\rightarrow -i[\,,\,]$.
In the case of the Dirac bracket \eqref{Diracbrktsb},
there is an obvious ordering problem. The choice of ordering
given below is the only one that is compatible with the supersymmetry 
algebra at the quantum level, eq. \eqref{qsymalg}, as will be shown in the next section. 
In this way, we obtain
\begin{subequations}\label{qmcomms}
\begin{align}
[x^a,\pi_b]&=i\delta^a{}_b, 
\label{qmcommsa}\\
[\pi_a,\pi_b]&=
-\frac{1}{2}R_{+cdab}(x)(\overline{\chi}{}_+{}^c\chi_+{}^d-
\chi_+{}^d\overline{\chi}{}_+{}^c)
\label{qmcommsb}\\
&\hskip 3cm-\frac{1}{2}R_{-cdab}(x)(\overline{\chi}{}_-{}^c\chi_-{}^d-
\chi_-{}^d\overline{\chi}{}_-{}^c), 
\nonumber\\
[\pi_a,\chi_\pm{}^b]&=i\Gamma_{\pm}{}^b{}_{ac}(x)\chi_\pm{}^c,
\label{qmcommsc}\\
[\pi_a,\overline{\chi}{}_\pm{}^b]&=i\Gamma_{\pm}{}^b{}_{ac}(x)\overline{\chi}{}_\pm{}^c,
\label{qmcommsd}\\
[\chi_\pm{}^a,\overline{\chi}{}_\pm{}^b]&=\Lambda_\pm{}^{ab}(x),
\label{qmcommse}
\end{align}
\end{subequations}
all remaining commutators vanishing. The commutation relations 
are compatible with the Jacobi identities, as is easy to check.
As to covariance, under a change of target space coordinates 
$t^a\rightarrow t'^a$, the 
operators $x^a$, $\pi_a$, $\chi_\pm{}^a$, $\overline{\chi}{}_\pm{}^a$
behave as their classical counterparts. For the operator $\pi_a$, there is
again an ordering problem. It can be seen that the ordering of the coordinate
transformation relation of $\pi_a$ compatible with covariance is
\begin{equation}\label{pi'pi}
\pi'{}_a=\frac{\partial t^b}{\partial t'^a}(x)\pi_b.
\end{equation}
Assuming this, the commutation relations \eqref{qmcomms} are straightforwardly 
checked to be covariant, as required. 

As explained earlier, the above quantization prescription is manifestly
covariant but not canonical. When studying Hilbert space representations of the 
quantum operator algebra, it may be convenient to have a canonical 
quantization prescription. To construct this, let us define
\begin{subequations}\label{epsiot}
\begin{align}
\epsilon^a&=\frac{1}{2^\frac{1}{2}}\big(\chi_+{}^a+\overline{\chi}{}_+{}^a
+i\chi_-{}^a+i\overline{\chi}{}_-{}^a\big),
\\
\iota_a&=\frac{1}{2^\frac{1}{2}}g_{ab}(x)\big(\chi_+{}^b+\overline{\chi}{}_+{}^b
-i\chi_-{}^b-i\overline{\chi}{}_-{}^b\big),
\end{align}
\end{subequations}
and set \hphantom{xxxxxxxxxxxxxxxxxxxxxxxxxxxxxxxx}
\begin{equation}\label{pa}
p_a=\pi_a-i\Gamma^c{}_{ab}(x)\epsilon^b\iota_c-\frac{i}{4}H_{abc}(x)\epsilon^b\epsilon^c
-\frac{i}{4}H_a{}^{bc}(x)\iota_b\iota_c.
\end{equation}
Using \eqref{qmcomms}, it is straightforward to verify that $x^a$, $p_a$,
$\epsilon^a$, $\iota_a$ satisfy the quantum graded commutation relations
\begin{subequations}\label{epsiotpcoms}
\begin{align}
[x^a,p_b]&=i\delta^a{}_b,
\label{epsiotpcomsa}\\
[\epsilon^a,\iota_b]&=\delta^a{}_b,
\label{epsiotpcomsb}
\end{align}
\end{subequations}
all other commutators vanishing. Note that, in particular, $[p_a,p_b]=0$,
while $[\pi_a,\pi_b]\not=0$. Thus, unlike the $\pi_a$, the momenta $p_a$ are canonical, as required.

Another issue related to quantization is that of the Hermiticity properties
of the operators $x^a$, $\pi_a$, $\chi_\pm{}^a$, $\overline{\chi}{}_\pm{}^a$. 
However, this problem cannot be posed at the level of target space local 
coordinate representations of these operators in general. Hermiticity is essentially 
a target space global property, since the Hilbert space product involves an
integration over $M$. This is obvious also from the coordinate
transformation relation \eqref{pi'pi}, which are not compatible with a naive 
Hermiticity relation of the form $\pi_a{}^*=\pi_a$. Similar considerations
apply to the canonical operators $x^a$, $p_a$, $\epsilon^a$, $\iota_a$.

This completes the quantization of biHermitian supersymmetric quantum mechanics.
In the Kaehler case, similar results were obtained in \cite{Bellucci1,Bellucci2}.

\vfill\eject

\begin{small}
\section{   \bf The quantum symmetry algebra}
\label{quantsusy}
\end{small}

As shown in section \ref{sec:susyqm}, 
biHermitian supersymmetric quantum mechanics has a rich symmetry structure
which should be reproduced 
at the quantum level. This leads 
to the requirement that 
the graded commutation relations 
\begin{subequations}\label{qsymalg}
\begin{align}
[Q_\pm,Q_\pm]&=0, 
\label{qsymalga}\\
[Q_\pm,Q_\mp]&=0,
\label{qsymalgb}\\
[\overline{Q}{}_\pm,\overline{Q}{}_\pm]&=0,
\label{qsymalgc}\\
[\overline{Q}{}_\pm,\overline{Q}{}_\mp]&=0,
\label{qsymalgd}\\
[Q_\pm,\overline{Q}{}_\pm]&=H_{QM},
\label{qsymalge}\\
[Q_\pm,\overline{Q}{}_\mp]&=0,
\label{qsymalgf}\\
[F_V,Q_\pm]&=-Q_\pm,
\label{qsymalgg}\\
[F_A,Q_\pm]&=\mp Q_\pm,
\label{qsymalgh}\\
[F_V,\overline{Q}{}_\pm]&=+\overline{Q}_\pm,
\label{qsymalgi}\\
[F_A,\overline{Q}{}_\pm]&=\pm \overline{Q}_\pm,
\label{qsymalgj}\\
[F_V,F_A]&=0
\label{qsymalgk}
\end{align}
\end{subequations}
should hold upon quantization. This symmetry algebra is in fact the quantum counterpart 
of the algebra \eqref{clsymalg} obeyed by the classical variation operators
$q_\pm$, $\overline{q}_\pm$, $f_V$, $f_A$ defined in \eqref{qmvarspr},
\eqref{Rqmvarspr}. Demanding that the operator relations
\eqref{qsymalg} hold is a very stringent requirement. It is not obvious a
priori that a quantization of the theory compatible with  
\eqref{qsymalg} exists, but in fact it does and it is unique. 
Indeed, imposing \eqref{qsymalg} determines not only all ordering ambiguities 
of the commutators of the basic operator variables $x^a$, $\pi_a$, 
$\chi_\pm{}^a$, $\overline{\chi}{}_\pm{}^a$, eqs. \eqref{qmcomms},
as anticipated in the previous section,
but also those of the supercharges $Q_{\pm}$, $\overline{Q}{}_\pm$ and the
Hamiltonian $H_{QM}$.  It does not determine conversely the ordering ambiguities 
of the vector and axial $R$ charges $F_V$, $F_A$. However, these ambiguities amount to a
harmless additive $c$ number that can be fixed conventionally, as can be
easily checked.

In this way, upon quantization, one finds that  $Q_{\pm}$, $\overline{Q}{}_\pm$ are given by
\begin{subequations}\label{qsusych}
\begin{align}
Q_\pm&=\chi_\pm{}^a\pi_a
\mp\frac{i}{2}H_{abc}(x)\chi_\pm{}^a\chi_\pm{}^b\overline{\chi}{}_\pm{}^c
\pm\frac{i}{2}H_{abc}\Lambda_\pm{}^{bc}(x)\chi_\pm{}^a,
\label{qsusycha}\\
\overline{Q}{}_\pm&=\overline{\chi}{}_\pm{}^a\pi_a
\mp\frac{i}{2}H_{abc}(x)\overline{\chi}{}_\pm{}^a\overline{\chi}{}_\pm{}^b\chi_\pm{}^c
\pm\frac{i}{2}H_{abc}\overline{\Lambda}{}_\pm{}^{bc}(x)\overline{\chi}{}_\pm{}^a.
\label{qsusychb}
\end{align}
\end{subequations}
The last term in the right hand side of both relations is a quantum ordering 
effect. Likewise, the quantum Hamiltonian is
\begin{align}\label{qH}
H_{QM}&=\frac{1}{2}\pi_ag^{ab}(x)\pi_b-\frac{i}{2}\Gamma^a{}_{ac}g^{cb}(x)\pi_b\\
&~~-\frac{1}{4}R_{+abcd}(x)(\overline{\chi}{}_+{}^a\chi_+{}^b-
\chi_+{}^b\overline{\chi}{}_+{}^a)(\overline{\chi}{}_-{}^c\chi_-{}^d-
\chi_-{}^d\overline{\chi}{}_-{}^c)
\nonumber\\
&~~+\frac{1}{8}R(x)-\frac{1}{96}H^{abc}H_{abc}(x).
\nonumber
\end{align}
Here, $\Gamma^a{}_{bc}$ is the usual Levi Civita connection and $R$ is the 
corresponding Ricci scalar. The second, fourth and fifth term in the right hand side
are again ordering effects.
Finally, the $R$ charges are given by
\begin{subequations}\label{qRch}
\begin{align}
F_V&=g_{ab}(x)\big(\overline{\chi}{}_+{}^a\chi_+{}^b+\overline{\chi}{}_-{}^a\chi_-{}^b\big)-\frac{n}{2},
\label{qRcha}\\
F_A&=g_{ab}(x)\big(\overline{\chi}{}_+{}^a\chi_+{}^b-\overline{\chi}{}_-{}^a\chi_-{}^b\big),
\label{qRchb}
\end{align}
\end{subequations}
where $n=\dim_\mathbb{R}M$, with a conventional choice of operator ordering. 

One can express the supercharges $Q_{\pm}$,
$\overline{Q}{}_\pm$, the Hamiltonian $H_{QM}$ and the $R$ charges $F_V$, $F_A$
in terms of the canonical operators $x^a$, $p_a$, $\epsilon^a$, $\iota_a$.
From \eqref{qsusych}, the supercharges are given by
\begin{subequations}\label{Qpmform}
\begin{align}
Q_\pm&=\Lambda_\pm{}^b{}_a(x)\psi_\pm{}^aP_{\mp b}
\pm\frac{i}{2}H_{abc}\Lambda_\pm{}^{bc}(x)\psi_\pm{}^a
\pm\frac{i}{6}H_{abc}(x)\psi_\pm{}^a\psi_\pm{}^b\psi_\pm{}^c,
\label{Qpmforma}\\
\overline{Q}{}_\pm&=\overline{\Lambda}{}_\pm{}^b{}_a(x)\psi_\pm{}^aP_{\mp b}
\pm\frac{i}{2}H_{abc}\overline{\Lambda}{}_\pm{}^{bc}(x)\psi_\pm{}^a
\pm\frac{i}{6}H_{abc}(x)\psi_\pm{}^a\psi_\pm{}^b\psi_\pm{}^c,
\label{Qpmformb}
\end{align}
\end{subequations}
where $\psi_\pm{}^a$, $P_{\pm a}$ are given by
\begin{subequations}\label{epsiotpinv}
\begin{align}
\psi_+{}^a&=\frac{1}{2^\frac{1}{2}}\big(\epsilon^a+g^{ab}(x)\iota_b\big),
\\
\psi_-{}^a&=\frac{1}{2^\frac{1}{2}i}\big(\epsilon^a-g^{ab}(x)\iota_b\big)
\end{align}
\end{subequations}
and \hphantom{xxxxxxxxxxxxxxxxxxxxxxxxxxxxxxxxxxxx}
\begin{equation}\label{Ppma}
P_{\pm a}=p_a+i\Gamma_\pm{}^c{}_{ab}(x)\epsilon^b\iota_c.
\end{equation}
From \eqref{qH}, the quantum Hamiltonian is 
\begin{align}
H_{QM}&=\frac{1}{2}g^{ab}(x)\pi_a\pi_b+\frac{i}{2}\Gamma^a{}_{bc}g^{bc}(x)\pi_a
-\frac{1}{4}R_{+abcd}(x)\psi_+{}^a\psi_+{}^b\psi_-{}^c\psi_-{}^d
\\
&~~+\frac{1}{8}R(x)-\frac{1}{96}H^{abc}H_{abc}(x),
\nonumber
\end{align}
where $\pi_a$ is given by 
\begin{equation}\label{pia}
\pi_a=p_a+i\Gamma^c{}_{ab}(x)\epsilon^b\iota_c+\frac{i}{4}H_{abc}(x)\epsilon^b\epsilon^c
+\frac{i}{4}H_a{}^{bc}(x)\iota_b\iota_c.
\end{equation}
Finally, from \eqref{qRch}, the $R$ charges are given by
\begin{subequations}\label{FVFAform}
\begin{align}
F_V&=\Lambda_{+ab}(x)\psi_+{}^a\psi_+{}^b
+\Lambda_{-ab}(x)\psi_-{}^a\psi_-{}^b-\frac{n}{2},
\\
F_A&=\Lambda_{+ab}(x)\psi_+{}^a\psi_+{}^b
-\Lambda_{-ab}(x)\psi_-{}^a\psi_-{}^b.
\end{align}
\end{subequations}

In this way, we succeeded in quantizing biHermitian supersymmetric quantum
mechanics in a way compatible with supersymmetry and $R$ symmetry. 
The next step of our analysis would be the search for interesting Hilbert 
space representations of the operator algebra. We shall postpone this to the next section.
Here, we shall analyze which properties a representation should have on
general physical grounds. 

A representation of the operator algebra in a Hilbert space $\mathcal{V}$
allows one to define the adjoint of any operator. The Hamiltonian $H_{QM}$ 
of biHermitian super\-symmetric quantum mechanics should be selfadjoint
\begin{equation}\label{H=Hdagger}
H_{QM}{}^*=H_{QM},
\end{equation}
to have a real energy spectrum.
From \eqref{H=Hdagger}, on account of \eqref{qsymalge}, the supercharges 
$Q_\pm$, $\overline{Q}_\pm$ should satisfy \hphantom{xxxxxxxxxxxxxxxxxxxx}
\begin{equation}\label{Qbar=Qdagger}
Q_\pm{}^*=\overline{Q}_\pm
\end{equation}
under adjunction.
The $R$ charges should also be selfadjoint
\begin{subequations}\label{FX=FXdagger}
\begin{align}
F_V{}^*&=F_V,
\label{FX=FXdaggera}\\
F_A{}^*&=F_A,
\label{FX=FXdaggerb}
\end{align}
\end{subequations}
to have a real $R$ charge spectrum.

From \eqref{qsymalgk}, \eqref{FX=FXdagger}, it follows that that the $R$
charges $F_V$, $F_A$ can simultaneously be diagonalized in $\mathcal{V}$. 
Since $F_V$, $F_A$ are the infinitesimal generators of the $U(1)_V$, $U(1)_A$
symmetry groups, they should satisfy the conditions
\begin{subequations}\label{exp2piF=1}
\begin{align}
\exp(2\pi iF_V)&=1,
\label{exp2piF=1a}\\
\exp(2\pi iF_A)&=1.
\label{exp2piF=1b}
\end{align}
\end{subequations}
This implies that the spectra of $F_V$, $F_A$ are subsets of $\mathbb{Z}$. 
It follows that the Hilbert space $\mathcal{V}$ has a direct sum 
decomposition of the form
\begin{equation}\label{dirctsum}
\mathcal{V}=\bigoplus_{k_V,k_A\in\mathbb{Z}}\mathcal{V}_{k_V,k_A},
\end{equation}
where $\mathcal{V}_{k_V,k_A}$ is the joint eigenspace of $F_V$, $F_A$ 
of eigenvalues $k_V$, $k_A$.

Only a finite number of the subspaces $\mathcal{V}_{k_V,k_A}$ are non zero.
A technical analysis outlined in appendix \ref{sec:appendixFVFA} shows that
$\mathcal{V}_{k_V,k_A}=0$ unless the conditions
\begin{subequations}\label{speckvka}
\begin{align}
-\frac{n}{2}\leq &k_V \pm k_A\leq\frac{n}{2},
\label{speckvkaa}
\\
&k_V\pm k_A=\frac{n}{2}\quad \text{mod $2$}
\label{speckvkab}
\end{align}
\end{subequations}  
are simultaneously satisfied. The \eqref{speckvka} imply that 
\begin{equation}\label{1speckvka}
-\frac{n}{2}\leq k_V,k_A\leq\frac{n}{2}
\end{equation} 
and further, that
\begin{subequations}\label{2speckvka}
\begin{align}
&-\frac{n}{2}+|k_V|\leq k_A  \leq\frac{n}{2}-|k_V|,
\label{2speckvkaa}\\
&-\frac{n}{2}+|k_A|\leq k_V  \leq\frac{n}{2}-|k_A|,
\label{2speckvkab}
\end{align}
\end{subequations} 
as is easy to see. Moreover, for fixed $k_V$ ($k_A$), two consecutive eigenvalues $k_A$
($k_V$) differ precisely by $2$ units. Thus, the non vanishing $\mathcal{V}_{k_V,k_A}$ 
can be arranged in a diamond shaped array as follows
\begin{equation}\label{diamond}
\begin{array}{ccccccc}
  &  &  &  \mathcal{V}_{0,n/2}&  &  &  \\
   &  & \cdots &  & \cdots &  &  \\
   &\mathcal{V}_{-n/2+1,1}&  &  &  &\mathcal{V}_{n/2-1,1}  &  \\
  \mathcal{V}_{-n/2,0} &  &  & \cdots &  &  & \mathcal{V}_{n/2,0} \\
   & \mathcal{V}_{-n/2+1,-1} &  &  &  &\mathcal{V}_{n/2-1,-1}  &  \\
   &  & \cdots & & \cdots &  &  \\
   &  &  &\mathcal{V}_{0,-n/2}  &  &  &  \\
\end{array}.
\end{equation} 
Using the the commutation relations \eqref{qsymalgg}--\eqref{qsymalgj}, it is
readily verified that the supercharges 
$Q_\pm$, $\overline{Q}{}_\pm$ act as follows
\begin{equation}\label{Qpmactdiag}
\xymatrix{
\mathcal{V}_{k_V-1,k_A+1} & &\mathcal{V}_{k_V+1,k_A+1} \\
 &\mathcal{V}_{k_V,k_A}
\ar[lu]^{Q_-}\ar[ru]^{\overline{Q}{}_+}
\ar[ld]^{Q_+}\ar[rd]^{\overline{Q}{}_-} & ~~~~~~~~~~~~~~~~~.\\
 \mathcal{V}_{k_V-1,k_A-1}& & \mathcal{V}_{k_V+1,k_A-1}
}
\end{equation}

From \eqref{qsymalge}, \eqref{Qpmactdiag}, it follows easily that 
the spaces $\mathcal{V}_{k_V,k_A}$ are invariant under
the Hamiltonian $H_{QM}$. Each space $\mathcal{V}_{k_V,k_A}$ contains a subspace
$\mathcal{V}^{(0)}{}_{k_V,k_A}$ spanned by the zero energy states 
$|\alpha\rangle\in\mathcal{V}_{k_V,k_A}$,
\begin{equation}
H_{QM}|\alpha\rangle=0.
\end{equation}
By \eqref{qsymalge}, these are precisely the supersymmetric states 
$|\alpha\rangle\in\mathcal{V}_{k_V,k_A}$,
\begin{equation}
Q_\pm|\alpha\rangle=\overline{Q}{}_\pm|\alpha\rangle=0.
\end{equation}

By \eqref{qsymalga}, \eqref{qsymalgc}, the supercharges 
$Q_\pm$, $\overline{Q}{}_\pm$ are nilpotent and, so, are characterized by
the cohomology spaces 
$H^{k_V,k_A}(Q_\pm)
=\ker Q_\pm\cap\mathcal{V}_{k_V,k_A}/ \im \,Q_\pm \cap\mathcal{V}_{k_V,k_A}$, 
$H^{k_V,k_A}(\overline{Q}{}_\pm)
=\ker \overline{Q}{}_\pm\cap\mathcal{V}_{k_V,k_A}/ \im \,\overline{Q}{}_\pm\cap\mathcal{V}_{k_V,k_A}$. 
By standard arguments of supersymmetric quantum mechanics, 
the $H^{k_V,k_A}(Q_\pm)$, $H^{k_V,k_A}(\overline{Q}{}_\pm)$
are all isomorphic to $\mathcal{V}^{(0)}{}_{k_V,k_A}$: 
each  state of $\mathcal{V}^{(0)}{}_{k_V,k_A}$
represents a distinct cohomology class of $Q_\pm$, $\overline{Q}{}_\pm$ and
each cohomology class of $Q_\pm$, $\overline{Q}{}_\pm$ has a unique 
representative state in $\mathcal{V}^{(0)}{}_{k_V,k_A}$.

Consider the total supercharge
\begin{equation}\label{Q}
Q=Q_++\overline{Q}{}_++iQ_-+i\overline{Q}{}_-
\end{equation}
and its adjoint \hphantom{xxxxxxxxxxxxxxxx}
\begin{equation}\label{Q*}
Q^*=Q_++\overline{Q}{}_+-iQ_--i\overline{Q}{}_-.
\end{equation}
$Q$, $Q^*$ satisfy the graded commutation relations
\begin{subequations}\label{Qalg}
\begin{align}
[Q,Q]&=0, 
\label{Qalga}\\
[Q^*,Q^*]&=0,
\label{Qalgb}\\
[Q,Q^*]&=4H_{QM},
\label{Qalgc}
\end{align}
\end{subequations}
It follows from here that the supercharges $Q$, $Q^*$ are nilpotent and that
their cohomology spaces $H^{k_V,k_A}(Q)$, $H^{k_V,k_A}(Q^*)$ are both isomorphic
to $\mathcal{V}^{(0)}{}_{k_V,k_A}$. 
Thus, the common cohomology of the supercharges
$Q_\pm$, $\overline{Q}{}_\pm$ can be described in terms of the total
supercharges $Q$, $Q^*$. This description is more transparent. Indeed, using 
\eqref{Qpmform}, \eqref {epsiotpinv}, \eqref{Ppma}, it is straightforward to verify that 
$Q$, $Q^*$ are given by 
\begin{subequations}\label{QQ*}
\begin{align}
Q&=-2^\frac{1}{2}i\big[\epsilon^aip_a-\frac{1}{6}H_{abc}(x)\epsilon^a\epsilon^b\epsilon^c\big],
\label{QQ*a}\\
Q^*&=+2^\frac{1}{2}i\big[-g^{ab}(x)\iota_a\big(ip_b-\Gamma^d{}_{bc}(x)\epsilon^c\iota_d\big)
+\frac{1}{6}H^{abc}(x)\iota_a\iota_b\iota_c\big].
\label{QQ*b}
\end{align}
\end{subequations}
From here, it appears that $Q$ depends on the $3$--form $H$ but it 
does not depend on metric $g$ and the complex structures
$K^\pm$ for given $H$. So does its cohomology.
This fact was noticed long ago in reference \cite{Rohm1}.

\vfill\eject

\begin{small}
\section{   \bf The differential form representation}
\label{derhamrep}
\end{small}
As in Riemannian supersymmetric quantum mechanics, the quantum operator algebra
of biHermitian supersymmetric quantum mechanics has a representation by
operators acting on a space of inhomogeneous complex differential forms. 

We assume first that the target manifold $M$ is compact. This ensures convergence of 
integration over $M$. 
The anticommutator algebra \eqref{epsiotpcomsb} of the canonical fermion 
operators $\epsilon^a$, $\iota_a$  has the well--known form of a 
fermionic creation and annihilation algebra. Therefore, it admits 
a standard Fock space representation. The Fock vacuum $|0\rangle$ is defined 
as usual by \hphantom{xxxxxxxxxxxxxxxxxxxxxxx}
\begin{equation}\label{|0>}
\iota_a|0\rangle=0.
\end{equation}
The most general state vector is of the form
\begin{equation}
|\omega\rangle=\sum_{p=0}^n\frac{1}{p!}\omega^{(p)}{}_{a_1\ldots a_p}(x)
\epsilon^{a_1}\ldots\epsilon^{a_p}|0\rangle,
\end{equation}
where the $\omega^{(p)}\in \Omega^p(M)\otimes\mathbb{C}$ are arbitrary complex
$p$--forms. Therefore, there is a one--to--one correspondence between state vectors and
inhomogeneous differential forms. The vacuum itself corresponds to the
constant $0$--form $1$, 
\begin{equation}\label{|0>1}
|0\rangle:=1.
\end{equation}
In this way, the action of the operators $x^a$, $p_a$,
$\epsilon^a$, $\iota_a$ on the state vector $|\omega\rangle$
is represented by operators acting on the space of inhomogeneous complex 
differential forms $\mathcal{V}=\Omega^*(M)\otimes\mathbb{C}$ according to the prescription
\begin{subequations}\label{epsiotpreps}
\begin{align}
x^a&:=t^a\cdot,
\label{epsiotprepsa}\\
p_a&:=-i\partial/\partial t^a,
\label{epsiotprepsb}\\
\epsilon^a&:=dt^a\wedge,
\label{epsiotprepsc}\\
\iota_a&:=\iota_{\partial/\partial t^a},
\label{epsiotprepsd}
\end{align}
\end{subequations}
where $t^a$ is a local coordinate. 
This yields the differential form representation of biHermitian 
supersymmetric quantum mechanics. In this representation,
the supercharges $Q_{\pm}$, $\overline{Q}{}_\pm$ and the $R$ charges $F_V$, $F_A$
are given by first and zeroth order differential operators on $\mathcal{V}$, 
respectively.

In the differential form representation, the Hilbert
space product is given by the usual formula
\begin{equation}\label{braket}
\langle\alpha|\beta\rangle=\frac{1}{~\mathrm{vol}(M)}\int_M d^nt \,g^\frac{1}{2}
\sum_p\frac{1}{p!}\alpha^{(p)a_1\ldots a_p}\beta^{(p)}{}_{a_1\ldots a_p},
\end{equation}
with $\alpha,\beta\in\mathcal{V}$. 
This allows to define the adjoint of the relevant operators. 
It is straightforward to check that
the supercharges $Q_{\pm}$, $\overline{Q}{}_\pm$, the Hamiltonian $H_{QM}$ 
and the $R$ charges $F_V$, $F_A$ satisfy the adjunction relations
\eqref{Qbar=Qdagger}, \eqref{H=Hdagger}, \eqref{FX=FXdagger}, 
respectively, as required.

If $M$ is not compact, we can repeat the above construction with a few changes.
It is necessary to restrict to differential forms with compact support. 
The state space is therefore isomorphic to $\mathcal{V}_c=\Omega_c{}^*(M)\otimes\mathbb{C}$. 
The Hilbert space product is given again by  
\eqref{braket} with the prefactor $1/\mathrm{vol}(M)$ removed and 
$\alpha,\beta\in\mathcal{V}_c$. Note that the Fock vacuum $|0\rangle$ is no
longer normalized.

\vfill\eject

\begin{small}
\section{   \bf Relation to generalized Hodge theory}
\label{hodge}
\end{small}

The algebraic framework described in the second half of section \ref{quantsusy}
is very reminiscent of the Hodge theory of compact generalized Kaehler manifolds
developed by Gualtieri in \cite{Gualtieri1} and reviewed briefly below 
\cite{Cavalcanti,Cavalcanti1}.
Indeed, in the differential form representation of section \ref{derhamrep},
they coincide, as we show below. This result, 
besides being interesting in its own, sheds also light on the nature of 
the space of supersymmetric ground states of biHermitian supersymmetric
quantum mechanics. In what follows, we assume that $M$ is compact.

The bundle $TM \oplus T^*M$ is endowed with a canonical indefinite metric. 
The Clifford bundle $C\ell(TM \oplus T^*M)$ is thus defined. 
The space of spinor fields of $C\ell(TM \oplus T^*M)$ is precisely the space
$\mathcal{V}=\Omega^*(M)\otimes\mathbb{C}$. The Clifford action of a section $X+\xi$ 
of $TM \oplus T^*M$ is defined by  
\begin{equation}\label{Cliffact}
(X+\xi)\mathbf{\cdot}=X^a(x)\iota_a+\xi_a(x)\epsilon^a,
\end{equation}
where \eqref{epsiotpreps} holds.
The biHermitian data $(g,H,K_\pm)$ are codified in two commuting $H$-twisted 
generalized complex structures
\begin{equation}\label{J1/2}
 \mathcal{J}_{1/2} = \frac{1}{2}  
\bigg(
\begin{array}{ll}
K_+ \pm K_- & (K_+\mp K_-)g^{-1} \\
g(K_+\mp K_-)& - (K_+{}^t \pm K_-{}^t)
\end{array} 
\bigg).
\end{equation}
defining a $H$-twisted generalized Kaehler structure \cite{Gualtieri}.
$\mathcal{J}_{1/2}$ are sections of the bundle
$\mathfrak{so}(TM \oplus T^*M)$.
They act on $\mathcal{V}$ via the Clifford action,
\begin{align}\label{JnCliff}
\mathcal{J}_{1/2}\mathbf{\cdot}&=\frac{1}{2}\Big\{\frac{1}{2}(K_+\mp K_-)^{ab}(x)\iota_a\iota_b
+\frac{1}{2}(K_+\mp K_-)_{ab}(x)\epsilon^a\epsilon^b
\\
&\hskip 5cm +\frac{1}{2}(K_+\pm K_-)^a{}_b(x)(\iota_a\epsilon^b-\epsilon^b\iota_a)
\Big\}.
\nonumber
\end{align}
It can be shown that the operators $\mathcal{J}_{1/2}\mathbf{\cdot}$ commute and that their
spectra are subsets of $i\mathbb{Z}$ \cite{Gualtieri1}. In this way, $\mathcal{V}$ decomposes 
as a direct sum of joint eigenspaces $\mathcal{V}'{}_{k_1,k_2}$ 
of $\mathcal{J}_{1/2}\mathbf{\cdot}$ labeled by two integers 
$k_1,k_2\in\mathbb{Z}$,
\begin{equation}\label{dirctsumG}
\mathcal{V}=\bigoplus_{k_1,k_2\in\mathbb{Z}}\mathcal{V}'{}_{k_1,k_2},
\end{equation}
in analogy to \eqref{dirctsum}. 
Further, the non vanishing subspaces $\mathcal{V}'{}_{k_1,k_2}$
can be arranged in a diamond shaped array 
\pagebreak[2]
\begin{equation}\label{diamondG}
\begin{array}{ccccccc}
  &  &  &  \mathcal{V}'{}_{0,n/2}&  &  &  \\
   &  & \cdots &  & \cdots &  &  \\
   &\mathcal{V}'{}_{-n/2+1,1}&  &  &  &\mathcal{V}'{}_{n/2-1,1}  &  \\
  \mathcal{V}'{}_{-n/2,0} &  &  & \cdots &  &  & \mathcal{V}'{}_{n/2,0} \\
   & \mathcal{V}'{}_{-n/2+1,-1} &  &  &  &\mathcal{V}'{}_{n/2-1,-1}  &  \\
   &  & \cdots & & \cdots &  &  \\
   &  &  &\mathcal{V}'{}_{0,-n/2}  &  &  &  \\
\end{array}
\end{equation}
analogous to \eqref{diamond}.

The $H$ twisted differential $d_H=d-H\wedge$ is given by
\begin{equation}\label{dH}
d_H=\epsilon^aip_a-\frac{1}{6}H_{abc}(x)\epsilon^a\epsilon^b\epsilon^c,
\end{equation}
where, again, \eqref{epsiotpreps} holds. In \cite{Gualtieri1}, it is 
shown that $d_H:\mathcal{V}'{}_{k_1,k_2}\rightarrow
\mathcal{V}'{}_{k_1-1,k_2-1}\oplus\mathcal{V}'{}_{k_1-1,k_2+1}\oplus\mathcal{V}'{}_{k_1+1,k_2+1}\oplus
\mathcal{V}'{}_{k_1+1,k_2-1}$. Therefore, projecting on the
four direct summands, $d_H$ decomposes as a sum  of the form
\begin{equation}\label{dHexp}
d_H=\delta_++\delta_-+\overline{\delta}{}_++\overline{\delta}{}_-, 
\end{equation}
where the operators $\delta_\pm$, $\overline{\delta}{}_\pm$ act as 
\begin{equation}\label{QpmactdiagG}
\xymatrix{
\mathcal{V}'{}_{k_1-1,k_2+1} & &\mathcal{V}'{}_{k_1+1,k_2+1} \\
 &\mathcal{V}'{}_{k_1,k_2}
\ar[lu]^{\delta_-}\ar[ru]^{\overline{\delta}{}_+}
\ar[ld]^{\delta_+}\ar[rd]^{\overline{\delta}{}_-} & \\
 \mathcal{V}'{}_{k_1-1,k_2-1}& & \mathcal{V}'{}_{k_1+1,k_2-1}
}
\end{equation}
in a way analogous to \eqref{Qpmactdiag}.

Now, using \eqref{epsiotpinv}, \eqref{FVFAform}, \eqref{JnCliff}, it is straightforward to
verify that 
\begin{equation}\label{FX=Jn}
\mathcal{J}_1\mathbf{\cdot}=iF_V,\qquad \mathcal{J}_2\mathbf{\cdot}=iF_A.
\end{equation}
Thus, one has $\mathcal{V}'{}_{k_1,k_2}=\mathcal{V}_{k_V,k_A}$ for $k_1=k_V$, $k_2=k_A$.
So, the direct sum decomposition \eqref{dirctsum}, \eqref{dirctsumG}
of $\mathcal{V}$ coincide. 
From \eqref{QQ*a}, \eqref{dH}, it appears that 
\begin{equation}\label{QdHrel}
d_H=-\frac{1}{2^\frac{1}{2}i}Q
\end{equation}
where $Q$ is the total supercharge \eqref{Q}.   
Comparing \eqref{dHexp}, \eqref{Q}  and taking \eqref{QpmactdiagG},
\eqref{Qpmactdiag} into account leads immediately to the following identifications
\begin{subequations}\label{Qreldelta}
\begin{align}
\delta_+=-\frac{1}{2^\frac{1}{2}i}Q_+,
\\
\overline{\delta}{}_+=-\frac{1}{2^\frac{1}{2}i}\overline{Q}{}_+,
\\
\delta_-=-\frac{1}{2^\frac{1}{2}}Q_-,
\\
\overline{\delta}{}_-=-\frac{1}{2^\frac{1}{2}}\overline{Q}{}_-.
\end{align}
\end{subequations}
Combining \eqref{Qbar=Qdagger} and \eqref{Qreldelta}, one finds that
$\delta_\pm{}^*=\mp\overline{\delta}{}_\pm$, relations in fact obtained 
in \cite{Gualtieri1} by different methods.
Relations 
\eqref{Qreldelta} are the main 
results of this paper. \eqref{FX=Jn}, \eqref{QdHrel} were obtained in 
\cite{Kapustin2}. 

In \cite{Gualtieri1}, it is also shown that the Lie algebroid differentials
$\overline{\partial}_i$ associated to the generalized complex structures 
$\mathcal{J}_i$ are given by
\begin{subequations}\label{barpartial}
\begin{align}
\overline{\partial}_1&=\overline{\delta}{}_++\overline{\delta}{}_-,
\\
\overline{\partial}_2&=\overline{\delta}{}_++\delta_-.
\end{align}
\end{subequations}
From \eqref{Qreldelta}, one finds that the $\overline{\partial}_i$ are related
to the supercharges $Q_\pm$, $\overline{Q}{}_\pm$ as
\begin{subequations}\label{barpartialQrel}
\begin{align}
\overline{\partial}_1&=-\frac{1}{2^\frac{1}{2}i}Q_B,
\\
\overline{\partial}_2&=-\frac{1}{2^\frac{1}{2}i}Q_A,
\end{align}
\end{subequations}
where the supercharges $Q_B$, $Q_A$ are given by 
\begin{subequations}\label{QAB}
\begin{align}
Q_B&=\overline{Q}{}_++i\overline{Q}{}_-
\label{QABa}\\
Q_A&=\overline{Q}{}_++iQ_-
\label{QABb}
\end{align}
\end{subequations}
\footnote{$\vphantom{\Big[}$ The above definitions of $Q_A$, $Q_B$ differ from the 
conventional ones $Q_A=\overline{Q}{}_++Q_-$, $Q_B=\overline{Q}{}_++\overline{Q}{}_-$ 
by the factor $i$ multiplying the second term. This factor is due to the phase choice 
conventions of the supercharges $Q_\pm$, $\overline{Q}{}_\pm$ we adopted. Note that the algebra
\eqref{qsymalg} is invariant under the phase redefinitions $Q_\pm\rightarrow \mathrm{e}^{i\phi_\pm}Q_\pm$, 
$\overline{Q}{}_\pm\rightarrow \mathrm{e}^{-i\phi_\pm}\overline{Q}{}_\pm$. See
also section \ref{sec:amodel}.}.
In \cite{Kapustin2}, $Q_B$, $Q_A$ were related to the BRST charges
of the $B$, $A$ topological biHermitian sigma models.
We shall come back to this in section \ref{sec:amodel}.

The correspondence between the operators of Hodge theory of generalized
Kaehler geometry and those of biHermitian supersymmetric quantum mechanics is
summarized by overlapping 
the following diagrams
\begin{subequations}\label{hodgesusyrel}
\begin{align}
&\xymatrix{
\mathcal{V}_{k_1-1,k_2+1} & {}_{k_2}&\mathcal{V}_{k_1+1,k_2+1} \\
 &\mathcal{V}_{k_1,k_2}\ar@{.>}[r]^{\overline{\partial}{}_1}\ar@{.>}[l]^{\partial_1}
\ar@{.>}[u]^{\overline{\partial}{}_2}\ar@{.>}[d]^{\partial_2}
\ar[lu]^{\delta_-}\ar[ru]^{\overline{\delta}{}_+}
\ar[ld]^{\delta_+}\ar[rd]^{\overline{\delta}{}_-} &{}_{k_1} \\
 \mathcal{V}_{k_1-1,k_2-1}& & \mathcal{V}_{k_1+1,k_2-1}
}
\\
&\xymatrix{
\mathcal{V}_{k_V-1,k_A+1} & {}_{k_A}&\mathcal{V}_{k_V+1,k_A+1} \\
 &\mathcal{V}_{k_V,k_A}\ar@{.>}[r]^{Q_B}\ar@{.>}[l]^{\overline{Q}{}_B}
\ar@{.>}[u]^{Q_A}\ar@{.>}[d]^{\overline{Q}{}_A}\ar[lu]^{iQ_-}\ar[ru]^{\overline{Q}{}_+}
\ar[ld]^{Q_+}\ar[rd]^{i\overline{Q}{}_-} & {}_{k_V}\\
 \mathcal{V}_{k_V-1,k_A-1}& & \mathcal{V}_{k_V+1,k_A-1}
}
\end{align}
\end{subequations}
up to an overall factor $-1/2^\frac{1}{2}$. 

The $H$ twisted Laplacian is defined by 
\begin{equation}\label{DeltaH}
\Delta_H=[d_H,d_H{}^*].
\end{equation}
From \eqref{QdHrel}, \eqref{Qalgc}, it follows readily  that 
\begin{equation}\label{DeltaH=HQM}
\Delta_H=2H_{QM}. 
\end{equation}
Thus, the twisted Laplacian equals twice the quantum Hamiltonian. 

On view of \eqref{FX=Jn}, \eqref{QdHrel}, \eqref{DeltaH=HQM}, we see that the space of 
supersymmetric ground states of biHermitian supersymmetric quantum 
mechanics graded by the values of the vector and axial $R$ charges $F_V$,
$F_A$ can be identified with 
the complex $d_H$ cohomology $H_H{}^\bullet(M,\mathbb{C})$, or equivalently, with 
the space of complex $\Delta_H$ harmonic differential forms
$\mathrm{Harm}_H{}^\bullet(M,\mathbb{C})$ graded by 
$\mathcal{J}_1\mathbf{\cdot}$, $\mathcal{J}_2\mathbf{\cdot}$ eigenvalues. 
Such gradation constitutes the Hodge decomposition of the underlying twisted 
generalized Kaehler manifold $M$ as defined in \cite{Gualtieri1}:
\begin{equation}\label{hodgedecomp}
H_H{}^\bullet(M,\mathbb{C})=\mathrm{Harm}_H{}^\bullet(M,\mathbb{C})
=\bigoplus_{|k_1+k_2|\leq n/2,~k_1+
  k_2=n/2~\mathrm{mod}~2} \mathcal{V}^{(0)}{}_{k_1,k_2},
\end{equation}
where $\mathcal{V}^{(0)}{}_{k_1,k_2}$ are $\Delta_H$-harmonic forms in
$\mathcal{V}_{k_1,k_2}$.

In the usual Kaehler case, which occurs when $K_+=K_-=K$ and $H=0$,
the decomposition \eqref{hodgedecomp} differs from the familiar Dolbeault 
decomposition. In \cite{Michelsohn1}, Michelsohn called it 
the Clifford decomposition and showed that there is an
orthogonal transformation $u$, called the Hodge automorphism, taking
it to the usual Dolbeault decomposition. $u$ is given by
\begin{equation}\label{sl2c}
u=\exp\Big(\!-\frac{i\pi}{4} h\Big)\exp\Big(\frac{\pi}{4}(l^*-l)\Big),
\end{equation}
where $h$, $l$, $l^*$ are the generators of the
Kaehler $\mathfrak{sl}(2,\mathbb{C})$ symmetry algebra
\footnote{$\vphantom{\bigg]}$ $h$, $l$, $l^*$ are given explicitly by 
$h=n/2-\epsilon^a\iota_a$, $l=-(1/2)K^{ab}(x)\iota_a\iota_b$ 
$l^*=(1/2)K_{ab}(x)\epsilon^a\epsilon^b$ and satisfy the commutation relations 
$[h,l]=2l$, $[h,l^*]=-2l^*$, $[l,l^*]=h$.}. Indeed, it is straightforward to verify
that $u$ maps $\Omega^{p,q}(M)$ into $\mathcal{V}_{k_1,k_2}$ with
\begin{subequations}\label{k1k2pq}
\begin{align}
k_1&=-p+q,
\\
k_2&=+p+q-\frac{n}{2}.
\end{align}
\end{subequations}

In the Kaehler case, combining \eqref{Qreldelta},
\eqref{Qpmform}, one finds 
\begin{subequations}\label{Dolbeault}
\begin{align}
\delta_+=\frac{1}{2}(\partial-\overline{\partial}{}^*),
\\
\overline{\delta}{}_+=\frac{1}{2}(\overline{\partial}-\partial^*),
\\
\delta_-=\frac{1}{2}(\partial+\overline{\partial}{}^*),
\\
\overline{\delta}{}_-=\frac{1}{2}(\overline{\partial}+\partial^*),
\end{align}
\end{subequations}
where $\partial$, $\overline{\partial}$ are the Dolbeault differentials. From 
\eqref{barpartial}, one has then 
\begin{subequations}\label{barpartial1}
\begin{align}
\overline{\partial}_1&=\overline{\partial},
\label{barpartial1a}\\
\overline{\partial}_2&=\frac{1}{2}(d+id^{c*}),
\label{barpartial1b}
\end{align}
\end{subequations}
where $d=\overline{\partial}+\partial$ is the de Rham differential and 
$d^c=i(\overline{\partial}-\partial)$. 
Thus, $\overline{\partial}_1$ is nothing but the customary Dolbeault
differential $\overline{\partial}$, as expected. Conversely, 
$\overline{\partial}_2$ has no immediate simple interpretation, but, in fact, 
it is straightforwardly related to the de Rham differential $d$ by the Hodge
automorphism $u$ defined in \eqref{sl2c}:
\begin{equation}
\overline{\partial}_2=2^{-\frac{1}{2}}\mathrm{e}^{-\frac{i\pi}{4}}udu^{-1}.
\end{equation}

To conclude, we remark that the Hodge theory worked out in \cite{Gualtieri1}
is slightly more general than the one reviewed above, as it allows also 
for twisting by a field $B\in\Omega^2(M)$. $B$ should not be confused
with the field $b$ trivializing $H$ in \eqref{H=db}. $B$ acts on $\mathcal{V}$
via the Clifford action
\begin{equation}
B\mathbf{\cdot}=\frac{1}{2}B_{ab}(x)\epsilon^a\epsilon^b.
\end{equation}
$B$ twisting amounts to a similarity transformation 
\begin{equation}\label{Bsimtr}
X_B=\exp(B\mathbf{\cdot})X\exp(-B\mathbf{\cdot})
\end{equation}
of the relevant operators $X$ on $\mathcal{V}$. In particular, twisting transforms $d_H$ into
$(d_H)_B=d_{H+dB}$. From \eqref{Bsimtr}, it is clear that $B$ twisting
preserves the untwisted formal algebraic structure leaving the above
conclusions unchanged. However, the adjunction properties of the untwisted operators 
such as \eqref{Qbar=Qdagger}, \eqref{H=Hdagger}, \eqref{FX=FXdagger} 
are not enjoyed by their twisted versions, if one uses the Hilbert 
product defined in \eqref{braket}, but they do, if one also twists the 
product as follows
\begin{equation}\label{Bbracket}
\langle \alpha,\beta\rangle_B
=\langle \exp(-B\mathbf{\cdot})\alpha,\exp(-B\mathbf{\cdot})\beta\rangle/
\langle \exp(-B\mathbf{\cdot})1,\exp(-B\mathbf{\cdot})1\rangle,
\end{equation}
with $\alpha,\beta\in\mathcal{V}$. $\langle,\rangle_B$ is called the Born--Infeld 
product in \cite{Gualtieri1}, since
\begin{equation}
\langle\exp(-B\mathbf{\cdot})1,\exp(-B\mathbf{\cdot})1\rangle
=\frac{1}{~\mathrm{vol}(M)}\int_M d^nt \,g^{-\frac{1}{2}}\det(g+B).
\end{equation}
has the characteristic Born--Infeld form.

\vfill\eject

\begin{small}
\section{    \bf The biHermitian $A$ and $B$ sigma models}
\label{sec:amodel}
\end{small}

The topological twisting of the biHermitian $(2,2)$ supersymmetric sigma model 
is achieved by shifting the spin of fermions 
either by $F_V/2$ or $F_A/2$, where $F_V$, $F_A$ are the fermion's vector and axial
$R$ charges, respectively. The resulting topological sigma models will be
called biHermitian $A$ and $B$ models, respectively. 
As is well known, at the quantum level, the $R$ symmetries of the 
sigma model are spoiled by anomalies in general. 
The twisting can be performed only if the corresponding $R$ symmetry is non
anomalous. This happens provided the following conditions are satisfied \cite{Kapustin2}:
\begin{subequations}\label{ancanc1}
\begin{align}
&c_1(T_+^{1,0}M)-c_1(T_-^{1,0}M)=0,~~~~~~~\text{vector $R$  symmetry},
\label{ancanc1a}\\
&c_1(T_+^{1,0}M)+c_1(T_-^{1,0}M)=0,~~~~~~~\text{axial $R$  symmetry}.
\label{ancanc1b}
\end{align}
\end{subequations} 
$R$ symmetry anomaly cancellation, however, is not sufficient by itself 
to ensure the consistency of the
twisting. Requiring the absence of BRST anomalies
and the existence of a one--to--one state--operator correspondence  
implies further conditions
discussed in \cite{Kapustin2} in the framework of generalized Calabi--Yau
geometry.

To generate topological sigma models using twisting, 
we switch to the Euclidean version of the $(2,2)$ supersymmetric 
sigma model. Henceforth, $\Sigma$ is a compact Riemann surface of genus 
$\ell_\Sigma$. Further, the following formal substitutions are to be 
implemented  
\begin{subequations}
\begin{align}
\partial_{++}&\rightarrow\partial_z
\\
\partial_{--}&\rightarrow\overline{\partial}_{\overline{z}}
\\
\chi_+{}^a&\rightarrow\chi_{+\theta}{}^a\in 
C^\infty(\Sigma,\kappa_\Sigma{}^\frac{1}{2}\otimes x^*T_+^{1,0}M)
\\
\overline{\chi}_+{}^a&\rightarrow\overline{\chi}_{+\theta}{}^a\in C^\infty(\Sigma,
\kappa_\Sigma{}^\frac{1}{2}\otimes x^*T_+^{0,1}M)
\\
\chi_-{}^a&\rightarrow\chi_{-\overline{\theta}}{}^a
\in C^\infty(\Sigma,\overline{\kappa}_\Sigma{}^\frac{1}{2}\otimes x^*T_-^{1,0}M)
\\
\overline{\chi}_-{}^a&\rightarrow\overline{\chi}_{-\overline{\theta}}{}^a\in C^\infty(\Sigma,
\overline{\kappa}_\Sigma{}^\frac{1}{2}\otimes x^*T_-^{0,1}M)
\end{align}
\end{subequations}
where $\kappa_\Sigma{}^\frac{1}{2}$ is any chosen spin structure
(a square root of the canonical line bundle $\kappa_\Sigma$ of $\Sigma$).

The field content of the biHermitian $A$ sigma model is obtained 
from that of the $(2,2)$ supersymmetric sigma model via the substitutions
\begin{subequations}\label{Atwist}
\begin{align}
\chi_{+\theta}{}^a
&\rightarrow 
\psi_{+z}{}^a\in C^\infty(\Sigma,\kappa_\Sigma\otimes x^*T^{1,0}_+M),
\\
\overline{\chi}_{+\theta}{}^a
&\rightarrow
\overline{\chi}_+{}^a\in C^\infty(\Sigma,x^*T^{0,1}_+M),
\\
\chi_{-\overline{\theta}}{}^a
&\rightarrow
\chi_-{}^a\in C^\infty(\Sigma,x^*T^{1,0}_-M).
\\
\overline{\chi}_{-\overline{\theta}}{}^a
&\rightarrow
\overline{\psi}_{-\overline{z}}{}^a \in C^\infty(\Sigma,\overline{\kappa}_\Sigma\otimes x^*T^{0,1}_-M),
\end{align}
\end{subequations}
The BRST symmetry variations of the $A$ sigma model fields are obtained from those 
of the $(2,2)$ supersymmetric sigma model fields  (cf. eq. \eqref{22varspr}), by setting
\begin{subequations}\label{Atwistvars}
\begin{align}
\alpha^+=\tilde\alpha^-=0,
\label{Atwistvarsa}\\
\tilde\alpha^+=\alpha^-=\alpha.
\label{Atwistvarsb}
\end{align}
\end{subequations}
Similarly, the field content of the biHermitian $B$ sigma model is obtained 
from that of the $(2,2)$ supersymmetric sigma model via the substitutions
\begin{subequations}\label{Btwist}
\begin{align}
\chi_{+\theta}{}^a
&\rightarrow 
\psi_{+z}{}^a\in C^\infty(\Sigma,\kappa_\Sigma\otimes x^*T^{1,0}_+M),
\\
\overline{\chi}_{+\theta}{}^a
&\rightarrow
\overline{\chi}_+{}^a\in C^\infty(\Sigma,x^*T^{0,1}_+M),
\\
\chi_{-\overline{\theta}}{}^a
&\rightarrow
\psi_{-\overline{z}}{}^a \in C^\infty(\Sigma,\overline{\kappa}_\Sigma\otimes x^*T^{1,0}_-M),
\\
\overline{\chi}_{-\overline{\theta}}{}^a
&\rightarrow
\overline{\chi}_-{}^a\in C^\infty(\Sigma,x^*T^{0,1}_-M).
\end{align}
\end{subequations}
The BRST symmetry variations of the $B$ sigma model fields are obtained from those 
of the $(2,2)$ supersymmetric sigma model fields, by setting
\begin{subequations}\label{Btwistvars}
\begin{align}
\alpha^+=\alpha^-=0,
\label{Btwistvarsa}\\
\tilde\alpha^+=\tilde\alpha^-=\alpha.
\label{Btwistvarsb}
\end{align}
\end{subequations}
Inspection of the $A$, $B$ twist prescriptions reveals that
\begin{equation}\label{ABcomp}
\text{$A$ twist}~\leftrightarrows~\text{$B$ twist}  
~~\text{under} ~~K_-{}^a{}_b  \leftrightarrows -K_-{}^a{}_b.
\end{equation}
The target space geometrical data $(g,H,K_\pm)$, 
$(g,H_,\pm K_\pm)$ have precisely the same properties: they are both biHermitian
structures. So, at the classical level,
any statement concerning the $A$ ($B$)
model translates automatically into one concerning the $B$ ($A$) model
upon reversing the sign of $K_-$ 
\footnote{\vphantom{$\bigg[$} For notational consistency, 
exchanging $K_-{}^a{}_b  \leftrightarrows -K_-{}^a{}_b$ must be accompanied by 
switching $\alpha^-\leftrightarrows \tilde\alpha^-$.}. For this reason, below,
we shall consider only the $B$ twist, unless otherwise stated. 

The twisted action $S_t$ is obtained from the $(2,2)$ supersymmetric sigma
model action $S$ \eqref{22action} implementing the substitutions
\eqref{Btwist}. One finds \cite{Zucchini4}
\begin{align}
S_t&=\int_\Sigma d^2z\bigg[\frac{1}{2}(g_{ab}+b_{ab})(x)\partial_zx^a
\overline{\partial}_{\overline{z}}x^b
\label{topaction}\\
&\hskip1.9cm+ig_{ab}(x)(\psi_{+z}{}^a\overline{\nabla}_{+\overline{z}}\overline{\chi}_+{}^b
+\psi_{-\overline{z}}{}^a\nabla_{-z}\overline{\chi}{}_-{}^b)
\nonumber\\
&\hskip1.9cm
+R_{+abcd}(x)\overline{\chi}_+{}^a\psi_{+z}{}^b\overline{\chi}{}_-{}^c\psi_{-\overline{z}}{}^d
\bigg].
\nonumber
\end{align}
The topological field variations are obtained from the $(2,2)$
supersymmetry field variations \eqref{22varspr} via \eqref{Btwist}, \eqref{Btwistvars}.
In \eqref{Btwistvarsb}, there is no real need for the supersymmetry parameters
$\tilde\alpha^+$, $\tilde\alpha^-$ to take the same value $\alpha$, since,
under twisting both become scalars. If we insist $\tilde\alpha^+$,
$\tilde\alpha^-$ to be independent in \eqref{22vars}, we obtain a more general symmetry
variation
\begin{equation}\label{hatdeltatop}
\delta_t=\tilde\alpha^+s_{t+}+\tilde\alpha^-s_{t+}
\end{equation}
where the fermionic variation operators $s_{t\pm}$ are given by \cite{Zucchini4}
\begin{subequations}\label{stoppm}
\begin{align}
s_{t+}x^a&=i\overline{\chi}_+{}^a,
\\
s_{t-}x^a&=i\overline{\chi}{}_-{}^a,
\\
s_{\vphantom{f}t+}\overline{\chi}_+{}^a&=0,
\\
s_{\vphantom{f}t-}\overline{\chi}_+{}^a
&=-i\Gamma_+{}^a{}_{bc}(x)\overline{\chi}{}_-{}^b\overline{\chi}_+{}^c,
\\
s_{t+}\overline{\chi}_-{}^a&=-i\Gamma_-{}^a{}_{cb}(x)\overline{\chi}_+{}^c\overline{\chi}{}_-{}^b,
\\
s_{t-}\overline{\chi}_-{}^a&=0,
\end{align}
\pagebreak[2]
\begin{align}
s_{t+}\psi_{+z}{}^a&=
-i\Gamma_+{}^a{}_{bc}(x)\overline{\chi}_+{}^b\psi_{+z}{}^c
-\Lambda_+{}^a{}_b(x)(\partial_zx^b-iH^b{}_{cd}(x)\overline{\chi}_+{}^c\psi_{+z}{}^d),
\\
s_{t-}\psi_{+z}{}^a&=-i\Gamma_+{}^a{}_{bc}(x)\overline{\chi}_-{}^b\psi_{+z}{}^c,
\\
s_{t+}\psi_{-\overline{z}}{}^a&=
-i\Gamma_-{}^a{}_{bc}(x)\overline{\chi}_+{}^b\psi_{-\overline{z}}{}^c,
\\
s_{t-}\psi_{-\overline{z}}{}^a&=
-i\Gamma_-{}^a{}_{bc}(x)\overline{\chi}{}_-{}^b\psi_{-\overline{z}}{}^c
-\Lambda_-{}^a{}_b(x)(\overline{\partial}_{\overline{z}}x^b
+iH^b{}_{cd}(x)\overline{\chi}{}_-{}^c\psi_{-\overline{z}}{}^d).
\end{align}
\end{subequations}
The action $S_t$ is invariant under both $s_{t\pm}$ \cite{Zucchini4},
\begin{equation}\label{stoppmSt=0}
s_{t\pm}S_t=0.
\end{equation}
One can show also that the $s_{t\pm}$ are nilpotent and anticommute on shell
\begin{subequations}\label{stoppm2=0}
\begin{align}
&s_{t\pm}{}^2\approx 0,
\\
&s_{t+}s_{t-}+s_{t-}s_{t+}\approx 0,
\end{align}
\end{subequations}
where $\approx$ denotes equality on shell.
The remarks following eqs. \eqref{clsymalg} hold in this case as well. 
The usual topological BRST variation $s_t$ is obtained when 
\eqref{Btwistvarsb} is satisfied. $s_t$ and the $s_{t\pm}$ are related as 
\begin{equation}\label{st=st++st-}
s_t=s_{t+}+s_{t-}.
\end{equation}
\eqref{st=st++st-} corresponds to the decomposition of the BRST charge in its 
left and right chiral components. Clearly, one has   
\begin{equation}\label{st2=0}
s_t{}^2\approx 0.
\end{equation}
Therefore, the $s_{t\pm}$ define an on shell
cohomological double complex, whose total differential is $s_t$, a fact 
already noticed in \cite{Kapustin2}. The total on shell cohomology 
is isomorphic to the BRST cohomology.

Taking \eqref{Btwist} into account and comparing \eqref{stoppm} and 
\eqref{qmvarspr}, we see that, in the limit
in which the world sheet $\Sigma$ shrinks to a world line $T$, yielding 
the point particle approximation leading to biHermitian 
supersymmetric quantum mechanics, the sigma model variation 
operators $s_{t\pm}$ correspond to the quantum mechanics variation 
operators $\overline{q}{}_\pm$ and, so, in the full quantum theory,
to the supercharges $\overline{Q}{}_\pm$. Thus, the $s_{t\pm}$ 
correspond to the operators $\overline{\delta}{}_\pm$ of Gualtieri's
generalized Hodge theory via \eqref{Qreldelta}. This solves the problem of interpreting 
the $s_{t\pm}$ in the framework of generalized Kaehler geometry, which was posed
but not solved in \cite{Zucchini4}.

By \eqref{st=st++st-}, the topological BRST variation $s_t$ is the counterpart
of the supercharge $\overline{Q}{}_++\overline{Q}{}_-$. This is almost the 
supercharge $Q_B=\overline{Q}{}_++i\overline{Q}{}_-$ defined in \eqref{QABa}.
To turn it precisely in this form, we perform the phase redefinitions
$\overline{\chi}_-{}^a\rightarrow +i\overline{\chi}_-{}^a$, 
$\psi_{-\overline{z}}{}^a\rightarrow -i\psi_{-\overline{z}}{}^a$
and $s_{t-}\rightarrow +is_{t-}$. These redefintions leave both the action
$S_t$, eq. \eqref{topaction}, and the variation operators $s_{t\pm}$,
eqs. \eqref{stoppm}, invariant in form, as is easy to see, but lead to 
identifying $s_{t-}$ with $i\overline{Q}{}_-$ rather than $\overline{Q}{}_-$.
Upon doing this, $s_t$ corresponds to the supercharge $Q_B$ and thus to the 
Lie algebroid differential $\overline{\partial}{}_1$ of Gualtieri's theory via
\eqref{barpartialQrel}. Thus, we recover one of the main results of
reference \cite{Kapustin2}. 

Our analysis so far concerned the state BRST cohomology. 
One may also consider the operator BRST cohomology. 
In a topological field theory, the state and operator BRST 
complexes are isomorphic and, so, are their BRST cohomologies.
In \cite{Kapustin2}, it was shown that,
in order such correspondence to hold, the topological condition
\eqref{ancanc1b} is not sufficient (for the B model). 
It is necessary to require that $\mathcal{J}_1$ is a weak twisted generalized 
Calabi--Yau structure. Let us recall briefly the meaning of this notion. 

Let $E_1$ be the $-i$ eigenbundle of the twisted generalized
complex structure $\mathcal{J}_1$ in $(TM \oplus T^*M)\otimes \mathbb{C}$. 
With $E_1$, there is associated locally a nowhere vanishing
section $\rho_1$ of $\wedge^*T^*M\otimes\mathbb{C}$ defined up to pointwise 
normalization by the condition $s\cdot\rho_1=0$, for all sections
$s$ of $E_1$, where $\cdot$ denotes the Clifford action 
\eqref{Cliffact}. Globally, $\rho_1$ defines a line bundle $U_1$ in
$\wedge^*T^*M\otimes\mathbb{C}$, called the canonical line bundle of 
$\mathcal{J}_1$. By definition, $\mathcal{J}_1$ is a weak twisted generalized 
Calabi--Yau structure, if $U_1$ is twisted generalized holomorphically
trivial. This means that: $a$) $U_1$ is topologically trivial and, so, 
admits a global nowhere vanishing section $\rho_1$, which is a form in
$\Omega^*(M)\otimes\mathbb{C}$; $b$) $\rho_1$ is twisted generalized holomorphic,
\pagebreak[2]
\begin{equation}\label{partial1phi=0}
\overline{\partial}{}_1\rho_1=0.
\end{equation}
By part $a$, there exists a linear isomorphism
$\phi:C^\infty(M,\wedge^*\overline{E}{}_1)\rightarrow \Omega^*(M)\otimes\mathbb{C}$
defined by the relation \hphantom{xxxxxxxxxx}
\begin{equation}\label{stosrho1}
\phi(s)=s\cdot\rho_1,
\end{equation}
with $s$ a section of $\wedge^*\overline{E}{}_1$. By part $b$, $\phi$ has the property that 
\begin{equation}\label{chain1}
\overline{\partial}{}_1\phi(s)=\phi(\partial_{\overline{E}{}_1} s),
\end{equation}
where $\partial_{\overline{E}{}_1}$ is the Lie algebroid differential of the Lie
algebroid $\overline{E}{}_1$. Therefore, $\phi$ yields an isomorphism of the
differential complexes
$(C^\infty(M,\wedge^*\overline{E}{}_1),\partial_{\overline{E}{}_1})$
and $(\Omega^*(M)\otimes\mathbb{C},\overline{\partial}{}_1)$
and, so, of their cohomologies. 
The canonical line bundle $U_1$ is isomorphic to the determinant 
line bundle $\det E_1$, $U_1\simeq \det E_1$. 
The condition \eqref{ancanc1b} is equivalent to 
$c_1(E_1)=0$ and, thus, to the triviality of $\det E_1$. Therefore,
\eqref{ancanc1b} is equivalent only to part $a$ of the weak twisted generalized 
Calabi--Yau condition and, so, it implies the existence of an isomorphism
$\phi$ satisfying \eqref{stosrho1}, but not \eqref{chain1}. Part $b$ has the
further consequence \eqref{chain1}, that leads to aforementioned cohomology 
isomorphism. 

As shown in \cite{Kapustin2} and reviewed above, 
$(\Omega^*(M)\otimes\mathbb{C},\overline{\partial}{}_1)$
is just the state BRST complex $(\mathcal{V},Q_B)$. 
By the basic relation $(\partial_{\overline{E}{}_1}s)\cdot=[\overline{\partial}{}_1,s\cdot]$,
$(C^\infty(M,\wedge^*\overline{E}{}_1),\partial_{\overline{E}{}_1})$
may be identified with the operator BRST complex $(\mathcal{O},\widehat{Q}{}_B)$,
where $\mathcal{O}$ is the sigma model operator algebra and \hphantom{xxxxxxxxxxxxxxxxxxxxxxx}
\begin{equation}
\widehat{Q}{}_BO=[Q_B,O],
\end{equation}
with $O$ any operator. This identification hinges on the actual content of the
operator algebra $\mathcal{O}$, which we have not specified. 
Alternatively, one may use it as a definition of $\mathcal{O}$. 

Now, we know that we also have the complexes $(\Omega^*(M)\otimes\mathbb{C},\overline{\delta}{}_\pm)$
or, physically, the state complexes $(\mathcal{V},\overline{Q}{}_\pm)$. To
the $\overline{\delta}{}_\pm$, there should correspond nilpotent differentials $\delta_{\overline{E}{}_1\pm}$
in $C^\infty(M,\wedge^*\overline{E}{}_1)$ defined by 
\begin{equation}\label{chainpm}
\overline{\delta}{}_\pm\phi(s)=\phi(\delta_{\overline{E}{}_1\pm}s),
\end{equation}
with $s$ a section  of $\wedge^*\overline{E}{}_1$. Note that 
$(\delta_{\overline{E}{}_1\pm}s)\cdot=[\overline{\delta}{}_\pm,s\cdot]$. 
It is not obvious {\it a priori} that this really works, since the commutator in
the right hand side of this relation is in principle a first order differential operator, but it
actually does, as is easy to verify directly using the explicit expressions
of $\overline{\delta}{}_\pm$ obtainable from \eqref{Qreldelta}, \eqref{Qpmformb}.
In this way, we have obtained differential complexes 
$(C^\infty(M,\wedge^*\overline{E}{}_1),\delta_{\overline{E}{}_1\pm})$.
On the physical side, this should correspond to operator differential complexes
$(\mathcal{O},\widehat{\overline{Q}}{}_\pm)$, where $\mathcal{O}$ is as above
and the $\widehat{\overline{Q}}{}_\pm$ are defined by 
\begin{equation}
\widehat{\overline{Q}}{}_\pm O=[\overline{Q}{}_\pm,O],
\end{equation}
with $O$ any operator. 
 
The above considerations indicate that the variation 
operators $s_{t\pm}$ do not simply characterize the topological sigma model at the 
classical level, but have an operator counterpart $\overline{Q}{}_\pm$ 
at the quantum level. The cohomologies of $Q_B$, $\overline{Q}{}_\pm$ 
are pairwise isomorphic. Ultimately, the topological correlators are expected 
to depend only on their common cohomology.

\vfill\eject

\vskip.6cm\par\noindent{\bf Acknowledgments.}  
We thank F. Bastianelli and G. Velo for useful comments and constructive
criticism on the early versions of this work. 
\vfill\eject

\vfill\eject

\appendix

\begin{small}
\section{    \bf Formulae of biHermitian geometry}
\label{sec:appendixtens}
\end{small}
In this appendix, we collect a number of useful identities of biHermitian
geometry, which are repeatedly used in the calculations illustrated in the
main body of the paper. Below $(g, H,K_\pm)$
is a fixed biHermitian structure on an even dimensional manifold $M$. 

\begin{enumerate}
\item
Relations satisfied by the 3--form $H_{abc}$.
\begin{equation}
\partial_aH_{bcd}-\partial_bH_{acd}+\partial_cH_{abd}-\partial_dH_{abc}=0.
\end{equation}
 
\item
Relations satisfied by the connections $\Gamma_\pm{}^a{}_{bc}$.
\begin{subequations}
\begin{align}
&\Gamma_\pm{}^a{}_{bc}=\Gamma^a{}_{bc}\pm\frac{1}{2}H^a{}_{bc},
\\
&\Gamma_\pm{}^a{}_{bc}=\Gamma_\mp{}^a{}_{cb},
\end{align}
\end{subequations}
where $\Gamma^a{}_{bc}$ is the Levi--Civita connection of the metric $g_{ab}$.

\item
Relations satisfied by the torsion $T_\pm{}^a{}_{bc}$ of $\Gamma_\pm{}^a{}_{bc}$.
\begin{subequations}
\begin{align}
&T_\pm{}^a{}_{bc}=\pm H^a{}_{bc},
\\
&T_\pm{}^a{}_{bc}=T_\mp{}^a{}_{cb}.
\end{align}
\end{subequations}
 
\item
Relations satisfied by the Riemann tensor $R_{\pm abcd}$ of $\Gamma_\pm{}^a{}_{bc}$.
\begin{subequations}
\begin{align}
&R_{\pm abcd}=R_{abcd}\pm\frac{1}{2}(\nabla_dH_{abc}-\nabla_cH_{abd})
+\frac{1}{4}(H^e{}_{ad}H_{ebc}-H^e{}_{ac}H_{ebd}),
\\
&R_{\pm abcd}=R_{\mp cdab},
\end{align}
\end{subequations}
where $R_{abcd}$ is the Riemann tensor of the metric $g_{ab}$.
\par
Bianchi identities. 
\begin{subequations}
\begin{align}
&R_{\pm abcd}+R_{\pm acdb}+R_{\pm adbc}
\mp(\nabla_{\pm b}H_{acd}+\nabla_{\pm c}H_{adb}+\nabla_{\pm d}H_{abc})
\\
&\hskip5cm +H^e{}_{ab}H_{ecd}+H^e{}_{ac}H_{edb}+H^e{}_{ad}H_{ebc}=0,
\nonumber\\
&\nabla_{\pm e}R_{\pm abcd}+\nabla_{\pm c}R_{\pm abde}+\nabla_{\pm d}R_{\pm abec}
\\
&\hskip3.5cm \pm(H^f{}_{ec}R_{\pm abfd}+H^f{}_{cd}R_{\pm abfe}+H^f{}_{de}R_{\pm abfc})=0.
\nonumber
\end{align}
\end{subequations}
Other identities 
\begin{subequations}
\begin{align}
&R_{\pm abcd}-R_{\pm cbad}=R_{\pm acbd}\pm\nabla_{\pm d}H_{abc},
\\
&R_{\pm abcd}-R_{\pm cbad}=R_{\mp acbd}\mp\nabla_{\mp b}H_{acd},
\\
&R_{\pm abcd}-R_{\mp abcd}=\pm\nabla_{\pm d}H_{abc}\mp\nabla_{\pm c}H_{dab}
\\
&\hskip4.5cm +H^e{}_{ac}H_{ebd}+H^e{}_{da}H_{ebc}-H^e{}_{ab}H_{ecd}.
\nonumber
\end{align}
\end{subequations}

\item
The complex structures $K_\pm{}^a{}_cK_\pm{}^c{}_b$.
\begin{equation}
K_\pm{}^a{}_cK_\pm{}^c{}_b=-\delta^a{}_b.
\end{equation}
Integrability
\begin{equation}
K_\pm{}^d{}_a\partial_dK_\pm{}^c{}_b-K_\pm{}^d{}_b\partial_dK_\pm{}^c{}_a
-K_\pm{}^c{}_d\partial_aK_\pm{}^d{}_b+K_\pm{}^c{}_d\partial_bK_\pm{}^d{}_a=0.
\end{equation}
Hermiticity \hphantom{xxxxxxxxxxxxxxxxx}
\begin{equation}
g_{cd}K_\pm{}^c{}_aK_\pm{}^d{}_b=g_{ab}.
\end{equation}
Kaehlerness with torsion \hphantom{xxxxxxxxxxxxxxxxx}
\begin{equation}
\nabla_{\pm a}K_\pm{}^b{}_c=0.
\end{equation}

\item 
Other properties.
\begin{equation}
H_{efg}\Lambda_\pm{}^e{}_a\Lambda_\pm{}^f{}_b\Lambda_\pm{}^g{}_c=0~~\text{and c. c.},
\end{equation}
\begin{equation}
R_{\pm efcd}\Lambda_\pm{}^e{}_a\Lambda_\pm{}^f{}_b=0~~\text{and c. c.},
\end{equation}
where \hphantom{xxxxxxxxxxxxxxxxx}
\begin{equation}
\Lambda_\pm{}^a{}_b=\frac{1}{2}(\delta^a{}_b-iK_\pm{}^a{}_b)~~\text{and c. c.}
\end{equation}

\end{enumerate}

\vfill\eject

\begin{small}
\section{    \bf The spectrum of $F_V$, $F_A$}
\label{sec:appendixFVFA}
\end{small}

Define the operators \hphantom{xxxxxxxxxxxxxxxxxx}
\begin{subequations}
\begin{align}
F_\pm&=g_{ab}(x)\overline{\chi}{}_\pm{}^a\chi_\pm{}^b,
\\
\tilde F_\pm&=g_{ab}(x)\chi_\pm{}^a\overline{\chi}{}_\pm{}^b.
\end{align}
\end{subequations}
These operators are selfadjoint and non negative
\begin{subequations}\label{Fpm}
\begin{align}
F_\pm&=F_\pm{}^*\geq 0, 
\\
\tilde F_\pm&=\tilde F_\pm{}^*\geq 0.
\end{align}
\end{subequations}
Further they satisfy the relation 
\begin{equation}
F_\pm+\tilde F_\pm=\frac{n}{2}
\end{equation}
From \eqref{qRch}, it is easy to see that  
\begin{subequations}\label{FVFAFpm}
\begin{align}
F_V&=F_++F_--\frac{n}{2}=-\tilde F_+-\tilde F_-+\frac{n}{2},
\\
F_A&=F_+-F_-=-\tilde F_++\tilde F_-.
\end{align}
\end{subequations}
From \eqref{Fpm}, \eqref{FVFAFpm}, it is immediate to see that the joint
$F_V$, $F_A$ eigenspaces $\mathcal{V}_{k_V,k_A}$ vanish 
unless the two conditions \eqref{speckvkaa} hold.

For fixed $k_V$, the spaces $\mathcal{V}_{k_V,k_A}$ are obtained 
from $\mathcal{V}_{k_V,-n/2+|k_V|}$ by applying operators with 
vanishing $F_V$ charge and positive $F_A$ charge. The only such operators are
linear combinations $\chi_-{}^a\overline{\chi}{}_+{}^b$ with $F_V$, $F_A$ neutral
coefficients or products of such operators. These operators increase the
eigenvalue of $F_A$ by multiples of $2$. A similar reasoning holds
interchanging the roles of $k_V$, $k_A$.

\vfill\eject

\end{document}